\documentclass[lettersize,journal]{IEEEtran}
\usepackage{amsmath,amsfonts}
\usepackage{algorithmic}
\usepackage{algorithm}
\usepackage{amssymb}
\usepackage{array}
\usepackage{textcomp}
\usepackage{stfloats}
\usepackage[version=4]{mhchem}
\usepackage{lipsum}  
\let\labelindent\relax
\usepackage{enumitem}
\usepackage{xcolor}
\usepackage{url}
\usepackage{verbatim}
\usepackage{pgfplots}
\usepackage{comment}
\usepackage{tikz}
\usetikzlibrary{calc}
\usepackage{graphicx}
\usepackage[font=small,labelfont=bf]{caption}
\usepackage{cite}
\usepackage{subfigure}
\DeclareMathOperator*{\argmax}{arg\,max}
\DeclareMathOperator*{\argmin}{arg\,min}
\DeclareMathOperator\erfc{erfc}

\newcommand{\p}{\mathrm{p}}

\newcommand{\E}{\mathrm{E}}

\newcommand{\Var}{\mathrm{Var}}

\newcommand{\Std}{\mathrm{Std}}

\newcommand{\bin}{\mathrm{Bin}}

\begin{document}

\title{Adaptive Molecular Communication Receivers with Tunable Ligand-Receptor Interactions}
\author{Murat Kuscu,~\IEEEmembership{Member,~IEEE}
       \thanks{The author is with the Nano/Bio/Physical Information and Communications Laboratory (CALICO Lab), Department of Electrical and Electronics Engineering, Koç University, Istanbul, Turkey (e-mail: mkuscu@ku.edu.tr).}
	   \thanks{This work was supported in part by the EU Horizon 2020 MSCA-IF under Grant \#101028935, and by The Scientific and Technological Research Council of Turkey (TUBITAK) under Grant \#120E301.}}        
    


\maketitle

\begin{abstract}
Molecular Communications (MC) underpins signaling in biological systems, enabling information transfer through biochemical molecules. The prospect of engineering this natural communication mechanism has inspired the Internet of Bio-Nano Things (IoBNT) applications, which rely on heterogeneous collaborative networks of natural and engineered biological devices, as well as artificial micro/nanomachines. A key attribute of natural MC systems is their adaptability, ensuring accurate information transmission in dynamic, time-varying biochemical environments. Therefore, integrating biological adaptation techniques into artificial MC networks, which are expected to operate in various biochemical environments, such as inside human body, is essential for robust and biocompatible IoBNT applications. This study explores the design of bio-inspired adaptive MC receivers capable of tuning their response functions for maintaining optimal detection performance in scenarios with time-varying received signals. The proposed receiver architectures are based on ligand-receptor interactions, with adaptivity achieved by modifying the sigmoidal-shaped ligand-receptor response curve in response to fluctuations in received signal statistics. The performance of these adaptive receivers is evaluated across a range of MC scenarios, including those with stochastic background interference, inter-symbol interference (ISI), and degrading enzymes, which involve time-varying scaling or shifting of received signals. Numerical results demonstrate the significant improvement in detection performance provided by adaptive receivers in dynamic MC scenarios. 
\end{abstract}


\section{Introduction}
\label{sec:introduction}
\IEEEPARstart{B}{iological} signaling is a vital process for all living systems, and it often involves the transmission and sensing of biochemical molecules encoding information. This so-called Molecular Communications (MC) is believed to have been optimized through evolution in terms of energy efficiency and reliability in complex and dynamic biological environments \cite{akan2016fundamentals}. This has inspired researchers to explore the opportunities for engineering MC towards creating artificial heterogeneous networks of `bio-nano things' (e.g., artificial and natural cells, biosensors, and micro/nano robots) in order to enable the Internet of Bio-Nano Things (IoBNT) applications, ranging from continuous health monitoring, smart drug delivery, to biofabrication and biocomputing \cite{kuscu2021internet, egan2023toward}. 

A key feature of natural cells, ranging from prokaryotes to eukaryotes, is their adaptation capability, which enables accurate sensing and reliable information transfer in constantly fluctuating and changing biological environments \cite{khammash2021perfect}. Many of the adaptation strategies adopted by natural cells involve the modulation of signal transduction through cell-surface receptors, i.e., ligand receptors, which act as an interface between the cell's interior and extracellular millieu. These receptors selectively and stochastically bind to target ligands, which convey environmental cues or transmit information from other biological entities \cite{tu2018adaptation}. The information is then transduced into a secondary messenger representation, ultimately leading to a cellular response through a cascade of transmembrane and intracellular reactions \cite{baker2006signal}. Therefore, optimization of the sensitivity, dynamic range, and selectivity of this cell-surface sensory system for the time-varying statistics of the environment through dynamic regulation of ligand receptors is crucial for accurate sensing and communication. 

Natural cells employ various dynamic receptor regulation methods, such as receptor phosphorilation/methylation \cite{xiao1999desensitization}, receptor clustering \cite{mello2007effects}, receptor cooperativity \cite{ortega2022rational}, heterotropic allostery \cite{olsman2016allosteric}, sequestration \cite{ricci2016using}, up/down-regulation of receptor abundances \cite{tecsileanu2019adaptation}, as well as increasing receptor diversity (bet-hedging) \cite{kamino2020adaptive}, among many others \cite{tu2018adaptation}. These regulations influence the receptor input-output behavior and modulate the sigmoidal response curve of the cell-surface sensory system. They can adjust the steepness of the response curve (response sensitivity), as well as the broadness and the location of the dynamic range of the response, which is centered around the equilibrium dissociation constant, $K_\mathrm{D}$, a quantitative measure of receptor binding affinity for target ligands \cite{ricci2016using}. For example, receptor clustering and cooperativity increase response curve steepness without altering its location, indicating a greater sensitivity and a narrower dynamic range \cite{komorowski2019limited}. In contrast, phosphorylation and heterotropic allostery modulate binding affinity, shifting the dynamic range \cite{xiao1999desensitization, olsman2016allosteric, ricci2016using}. Expressing new receptors with different binding affinities for the same target ligands transforms the response curve into multiple sigmoidals, each representing a distinct sensitive region \cite{getz2001receptor}, effectively extending the dynamic range. Additionally, cells are known to incorporate regulatory integral feedback for dynamic regulation, facilitated by their intracellular chemical reaction networks (CRNs) \cite{aoki2019universal}.

In the context of IoBNT, artificial MC networks are typically envisioned for use in complex, dynamic biochemical environments, such as inside human body, interacting continuously with co-existing biological systems \cite{akyildiz2015internet}. Therefore, it is crucial to integrate biological adaptation techniques into artificial MC systems. Various efforts have focused on developing adaptive MC techniques, such as adaptive modulation and detection schemes, to address intersymbol interference (ISI), which results from the substantial memory inherent in diffusion-based MC channels, leading to time-varying received signals \cite{kuscu2019transmitter}. For example, in \cite{damrath2016low}, the authors proposed an adaptive detection method adjusting the binary decision threshold based on the last decoded message, mitigating the effects of ISI. In \cite{chang2017adaptive}, adaptive detection schemes were introduced for mobile MC channels with time-varying channel impulse response (CIR), incorporating dynamic CIR estimation as feedback. Moreover, researchers proposed adaptive transmission methods that dynamically adjust molecule release rates to overcome ISI at the receiver \cite{movahednasab2015adaptive}. Although effective in addressing ISI, these techniques lack direct physical correspondence with the biological architectures of the MC transceivers. As such, the current research on adaptivity in engineered MC systems is limited to algorithmic or `software' solutions, and does not extend to the opportunities enabled by the flexible and dynamic nature of the physical architecture or `hardware' of the MC transceivers. 

In this study, we investigate adaptive biosynthetic MC receiver architectures capable of dynamically adjusting receiver response to maintain optimal detection performance in cases with time-varying received signal statistics. We consider time-varying diffusion-based MC channels, where information is encoded in the concentration of a single type of ligand. The receiver architectures under examination utilize cell-surface receptors that engage in reversible and monovalent interactions with information-carrying ligands. Specifically, we focus on two distinct adaptive MC receiver architectures that can either (i) tune $K_\mathrm{D}$ of receptors, thereby shifting the position of the dynamic range of the receiver response without changing its steepness or broadness, or (ii) express a second type of receptor on the cell surface, with a different affinity (i.e., different $K_\mathrm{D}$) for the ligands, modifying the response curve shape and expanding the dynamic range. 

To evaluate performance improvement enabled by adaptive MC receivers, we first examine two main types of potential variations in received signal statistics for time-varying MC channels: (i) scaling of received concentration signals, resulting from factors such as ligand degradation due to enzymes, or fluctuating/decreasing transmit power of MC transmitter, and (ii) shift of received concentration signals, which may arise from ISI or interference caused by background ligand concentrations. These variations can render received signals outside the dynamic range of receiver response, i.e., in insensitive regions such as the saturation region, where distinguishing between different symbols (e.g., bit-$0$ and bit-$1$) encoded in distinct ligand concentrations becomes challenging. This leads to suboptimal detection performance, which can be addressed by adaptive receivers with tunable ligand-receptor interactions.  

Next, we evaluate the performance of adaptive receivers in three distinct, practical, and time-varying MC scenarios extensively studied in the literature. The first scenario is the degradation of information-carrying ligands due to degrading enzymes present in the channel. The second scenario involves stochastic background interference, wherein background molecular concentration, which may result from an adjacent MC channel (i.e., co-channel interference) or a co-existing biological system, shifts the received signals. The third scenario highlights the effect of ISI, adding a time-varying and stochastic offset to the received signals. This scenario is also studied in the presence of a receiver memory that stores previously decoded bits for ISI estimation. Numerical results demonstrate significant error performance improvement in each case, facilitated by adaptive receiver architectures, and highlight the potential of combining algorithmic and physical architecture-based adaptivity. 

The main contributions of this paper can be summarized as follows:
\begin{itemize}[leftmargin=*]
\item Introduction of a simple, unifying analysis framework for time-varying MC scenarios of practical importance, highlighting the necessity of adaptivity.
\item Analysis of optimal $K_\mathrm{D}$ values and corresponding receptor response curves under various time-varying channel conditions, determining set points for optimal MC system performance.
\item Proposal of two biologically-relevant adaptive MC receiver architectures based on tunable ligand-receptor interactions.
\item Comprehensive analysis of communication performance improvement enabled by adaptivity in terms of Bit Error Probability (BEP), which can inform the design of robust, adaptive MC networks for practical IoBNT applications.
\end{itemize}
 
The problem, analysis, and results presented in this paper are particularly relevant for biosynthetic MC transceivers, which are envisioned based on engineered or artificial cells; and this discussion is timely given the recent advancements in synthetic biology. Significant progress has been made in rational design and engineering of synthetic receptors, combining natural biological components with tunable binding characteristics \cite{ricci2016using, ortega2022rational, chang2020synthetic}. This also includes \emph{de novo} design of proteins and receptors that can be integrated into biosynthetic cells \cite{huang2016coming}, as well as synthetic CRNs capable of arithmetic and logic operations, and regulatory functions, e.g., integral feedback \cite{frei2021adaptive}. Additionally, advances in electromagnetic modulation of protein interactions promise dynamic and external control in biosynthetic device architectures \cite{zimmerman2016tuning, elayan2020regulating}. Combining these emerging capabilities with an understanding of the adaptivity requirements of MC networks can contribute to developing physically relevant, robust, and biocompatible IoBNT applications \cite{egan2023toward}.

The remainder of this paper is organized as follows. In Section \ref{sec:system_model}, we present the MC channel and received signal models, and obtain optimal $K_\mathrm{D}$ for receptors. In Section \ref{sec:adaptive_receivers}, we introduce the adaptive MC receiver architectures, and provide a simple analysis framework for investigating their performance under two main types of time-varying received signal conditions. This analysis is extended to practical time-varying MC scenarios in Section \ref{sec:practical}, covering three sources of time-variance in MC channel: enzymatic degradation of ligands, ISI, and stochastic background interference. Concluding remarks are provided in Section \ref{sec:conclusion}. 

\section{MC System Model and Receiver Response}
\label{sec:system_model}

\subsection{MC Channel Model}

We consider MC between two biosynthetic cells, e.g., engineered (top-down designed) or artificial (bottom-up designed) cells, in a three-dimensional (3D) unbounded fluidic channel where molecules undergo free diffusion. One of the cells acts as a transmitter encoding information into the number of ligands it instantly releases into the channel at the beginning of a signaling interval following concentration shift keying (CSK) modulation \cite{kuscu2019transmitter}. The other cell acts as a receiver capable of detecting these molecular signals within the the same signaling interval through ligand receptors inside its lipid membrane. As per the convention in MC literature \cite{pierobon2011noise, kuscu2019channel}, we consider the lipid membrane of the receiver as a reception volume, in which membrane receptors are uniformly distributed at any given time. We also assume that transmitter and receiver cells are synchronized. 

Considering the distance between the transmitter and the receiver to be much larger than their sizes, we can assume that transmitter is a point source of molecules \cite{jamali2019channel}. On the same grounds, we also assume that the ligand concentration is uniformly distributed within the receiver's reception volume, equal to its value at the receiver's center position. With these assumptions, when the transmitter cell instantly releases a specific number of ligands, $N_\mathrm{L}$, encoding a symbol, the received ligand concentration signal as a function of time is given by 
\begin{equation}
c_\mathrm{L}(t) = h(t,d) N_\mathrm{L} + c_\mathrm{int}(t),
\label{eq:receivedsignal}
\end{equation}
where $h(t,d)$ is CIR of the free diffusion channel, which can be obtained by solving Fick's second law of diffusion: 
\begin{equation}
h(t,d) = (4 \pi D t)^{-3/2} \exp\left(-\frac{d^2}{4Dt}\right),
\label{eq:MCimpulseresponse}
\end{equation}
where $d$ is the distance between the transmitter and the receiver, and $D$ denotes the diffusion coefficient of ligands \cite{meng2014receiver}. In \eqref{eq:receivedsignal}, $c_\mathrm{int}(t)$ represents the interference caused by the concentration of same type of ligands that are not part of the intended signal transmission. This interference may originate from an unrelated MC system operating within the same channel or from a biological process \cite{dinc2017theoretical, kuscu2022detection}.

\subsection{MC Receiver with Ligand Receptors}

In natural cells, which serve as the basis for biosynthetic cells, received concentration signals external to the cell are translated into an intracellular representation through cell-surface receptors \cite{berg1977physics}. These receptors provide a high level of selectivity (or specificity) in sensing and communications within dynamic and complex biological environments \cite{bongrand1999ligand}. Typically, ligands bind reversibly to receptors, triggering a first response via activation or conformational change of receptors that is then conveyed intracellularly through secondary messengers. The stochastic binding state of the receptors, as reflected in the intracellular molecular composition, enables the cell to infer the characteristics of extracellular molecular signals in terms of ligand types and concentrations \cite{ten2016fundamental,singh2017simple}. 

In the case of monovalent receptors, which possess a single ligand-binding site and function independently of other membrane receptors and intracellular dynamics, reversible ligand-receptor interactions are represented by the following reaction: 
\begin{equation}
\ce{\mathrm{U}  <=>[{c_\mathrm{L}(t) k^+}][{k^-}] \mathrm{B}},
\label{eq:bindingreaction}
\end{equation}
where $\mathrm{U}$ and $\mathrm{B}$ indicate the unbound and bound states of the receptor, respectively, while $k^+$ and $k^-$ denote the binding and unbinding rates of the ligand-receptor pair, respectively \cite{ten2016fundamental}.

Due to the low-pass characteristics of the diffusion-based MC channel, the bandwidth of $c_\mathrm{L}(t)$ is typically significantly lower than the binding reaction's characteristic frequency, i.e., $f_B = c_\mathrm{L}(t) k^+ + k^-$ \cite{pierobon2011noise}. Therefore, the ligand-receptor reaction is generally assumed to be at equilibrium with a stationary ligand concentration, denoted simply as $c_\mathrm{L}$. At equilibrium, the probability of observing a receptor in the bound state as a function of ligand concentration is given as
\begin{equation}
\p_\mathrm{B}(c_\mathrm{L}) = \frac{c_\mathrm{L}}{c_\mathrm{L} + K_\mathrm{D}}.
\label{eq:probofbinding}
\end{equation}
The equilibrium dissociation constant, $K_\mathrm{D}$, in this equation is a crucial parameter that quantifies the affinity of the ligand-receptor pair through the expression: $K_\mathrm{D} = k^-/k^+$ \cite{bongrand1999ligand}. When multiple receptors are independently exposed to the same ligand concentration and do not interact with one another, the number of bound receptors becomes a binomial random variable with a success probability of $\p_\mathrm{B}$, i.e., $n_\mathrm{B} \sim \bin(\p_\mathrm{B},N_\mathrm{R})$, where $N_\mathrm{R}$ denotes the total number of receptors \cite{pierobon2011noise}. The mean and variance of the number of bound receptors can then be expressed as
\begin{align}
\E[n_\mathrm{B}] &= \p_\mathrm{B}(c_\mathrm{L}) N_\mathrm{R} \\ \nonumber
\Var[n_\mathrm{B}] &= \p_\mathrm{B}(c_\mathrm{L}) \bigl(1-\p_\mathrm{B}(c_\mathrm{L})\bigr) N_\mathrm{R}.
\label{eq:nBmeanvariance}
\end{align}
Through the relation $\p_\mathrm{B}(c_\mathrm{L}) = \E[n_\mathrm{B} / N_\mathrm{R}]$, we can also refer to $\p_\mathrm{B}(c_\mathrm{L})$ as \emph{the expected ratio of bound receptors} given the ligand concentration. For sufficiently high $N_\mathrm{R}$, $n_\mathrm{B}$ can be approximated as Gaussian distributed, i.e., $n_\mathrm{B} \sim \mathcal{N}(\E[n_\mathrm{B}],\Var[n_\mathrm{B}])$, although caution must be taken for very small and very high values of $\p_\mathrm{B}(c_\mathrm{L})$ \cite{kuscu2022detection}. For simplicity of our error analyses in the following sections, we use Gaussian approximation for $n_\mathrm{B}$. 

\subsection{Receiver Response}
\begin{figure}[!t]
	\centering
	\includegraphics[width=\columnwidth]{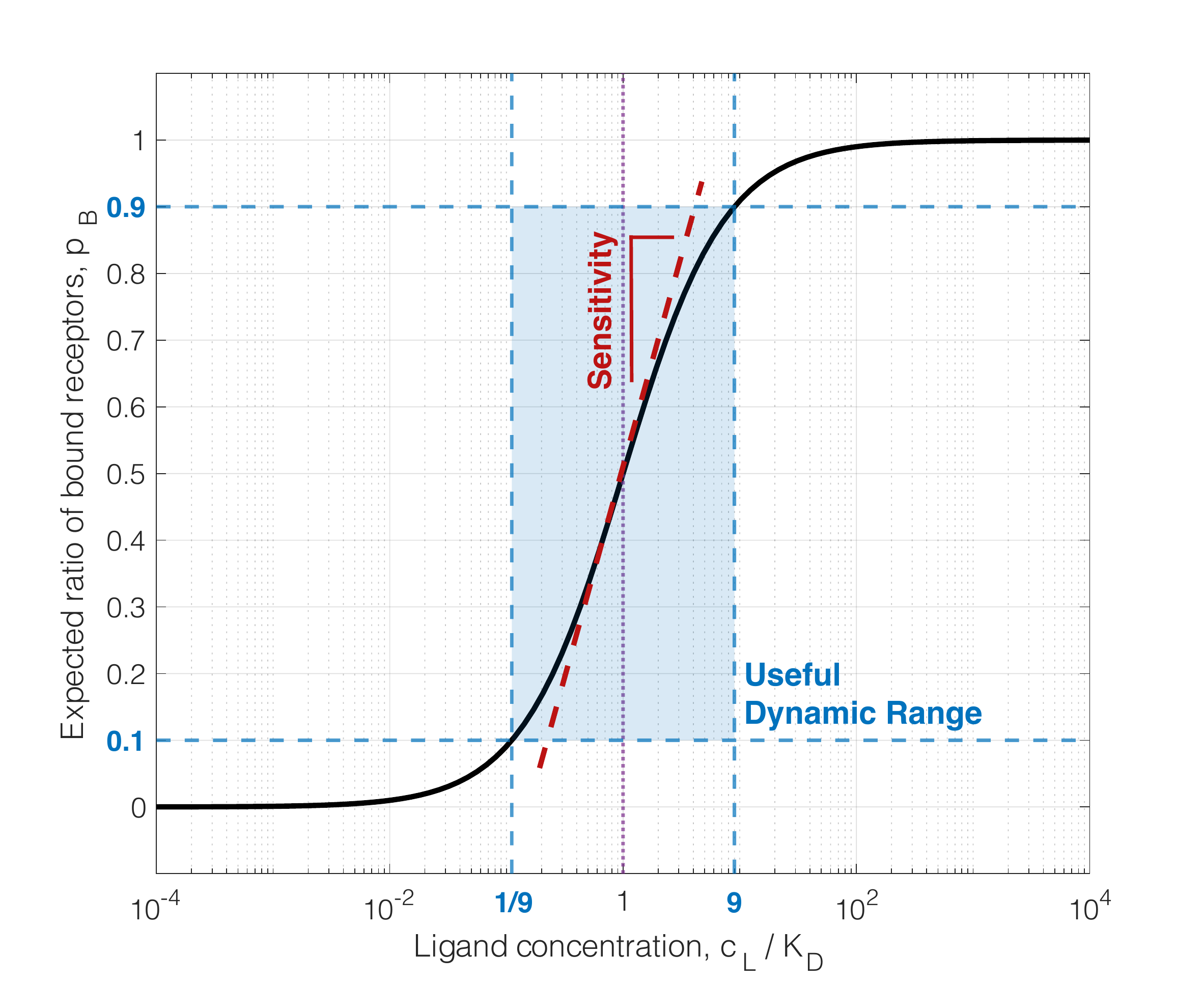}
	\caption{Sigmoidal response curve of the receiver with its sensitivity and useful dynamic range indicated on the curve.}
    \label{fig:responsecurve}
\end{figure}

In this study, we neglect intracellular dynamics, which may involve receptor activation and expression of secondary messengers, and assume that the receiver performs detection directly using \emph{the number of bound receptors}, $n_\mathrm{B}$, sampled at a given time instance within a signaling interval. Consequently, the receiver response curve can be represented by $\p_\mathrm{B}(c_\mathrm{L})$ and is plotted in Fig. \ref{fig:responsecurve} on a logarithmic scale. As indicated on the sigmoidal response curve, the curve's slope determines the \emph{sensitivity}, while the range of concentrations where the receiver is significantly responsive to small changes in ligand concentration defines its \emph{useful dynamic range}. This dynamic range is commonly represented as the range of concentrations corresponding to the $10-90\%$ expected ratio of bound receptors \cite{ricci2016using}. From Fig. \ref{fig:responsecurve} and \eqref{eq:probofbinding}, it becomes apparent  that both sensitivity and dynamic range are entirely dependent on the dissociation constant $K_\mathrm{D}$. 

The ability to adjust $K_\mathrm{D}$ for modulating the receiver response curve, thereby maintaining the sensitivity against variations in the received ligand concentration, lies at the heart of the adaptive receiver architectures investigated in this paper. The ligand concentration encoding capacity of receptors, which refers to the capability of unequivocally representing two distinct ligand concentration levels in terms of the number of bound receptors, is maximized when the logarithm of $K_\mathrm{D}$ is centered around the logarithmic concentration levels \cite{getz2001receptor}, i.e.,
\begin{equation}
\log(K_\mathrm{D,opt}) = \frac{1}{2} \left(\log(c_{\mathrm{L}|0}) + \log(c_{\mathrm{L}|1})\right),
\label{eq:optKDlog}
\end{equation}
or equivalently, when $K_\mathrm{D}$ is equal to the geometric mean of the concentration levels: 
\begin{equation}
K_\mathrm{D,opt} = \sqrt{c_\mathrm{L|0} ~c_\mathrm{L|1}}.
\label{eq:optKDlog}
\end{equation}
Here,  
\begin{equation}
c_{\mathrm{L}|s} = N_{\mathrm{L}|s} ~h(t_\mathrm{S}, d) + c_\mathrm{int}(t_\mathrm{S}),~~ \text{with}~ s \in \{0,1\},
\label{eq:receivedsignals}
\end{equation}
represents the received ligand concentration at the receiver at the sampling time $t_\mathrm{S}$, with $N_{L|s}$ denoting the transmitted number of ligands for encoding bit-$s \in \{0,1\}$, considering binary CSK (B-SCK) modulation is employed. In MC studies, the sampling time is typically set to the time point where the concentration of transmitted ligands reaches its maximum at the receiver, i.e., $t_\mathrm{S} = t_\mathrm{Peak}$ \cite{kuscu2019transmitter}. This peak time can be calculated from CIR given in \eqref{eq:MCimpulseresponse} as $t_\mathrm{Peak} = d^2/6D$. 

It can be demonstrated that $K_\mathrm{D,opt}$ is also optimal in terms of BEP, when the transmission probabilities of bit-$1$ and bit-$0$ are equal, i.e., $\p_1 = \p_0 = 1/2$, and the receiver performs detection using $n_\mathrm{B}$ (sampled at $t_\mathrm{Peak}$) and an optimum detection threshold. This can be shown by taking the derivative of the analytical BEP expression \eqref{eq:bepbep}, derived in Appendix \ref{AppendixA}, with respect to $K_\mathrm{D}$, and subsequently equating the derivative to zero and solving the resulting equation for $K_\mathrm{D}$ (derivation is not included). The adaptive receiver architectures explored in the following sections mainly build upon this fact.

\section{Adaptive MC Receivers under Received Signal Variations}
\label{sec:adaptive_receivers}

The MC scenarios considered in this paper share a common motif: The receiver is utilized in an MC application with a single type of receptors, which have a dissociation constant that is optimal with respect to the baseline (i.e., initial) setting or state of the MC system. In other words, the receptors have the optimal dissociation constant, $K_\mathrm{D,opt}^\ast = \sqrt{c_{\mathrm{L}|0}^\ast c_{\mathrm{L}|1}^\ast}$, where $c_{\mathrm{L}|s}^\ast = h^\ast(t_\mathrm{S},d^\ast) N_{\mathrm{L}|s}^\ast + c_\mathrm{int}^\ast$, with $^\ast$ denoting the baseline attributes of the MC system functions and parameters. The received signals subsequently undergo variations from their expected values $c_{\mathrm{L}|s}^\ast$, which may directly result from alterations in the form of CIR $h(\cdot)$, or in the value of the transmitter-receiver distance $d$, the transmitted number of ligands $N_{L|s}$, or the interference concentration $c_\mathrm{int}$. Consequently, the initial optimality of the (baseline) receptors $K_\mathrm{D,opt}^\ast$ is compromised. 

Regardless of the underlying cause, we will first examine two main types of received signal variations that may be encountered in practical applications: 
\begin{itemize}[leftmargin=*]
\item \textbf{Scaling of received concentration signals:} The received ligand concentrations corresponding to bit-$0$ and bit-$1$ can be scaled by the same scaling factor. Example scenarios where this may occur include practical MC systems with channels that feature enzymes degrading the information molecules \cite{noel2014improving}, or MC systems where the transmitter experiences diminishing transmit power due to the depletion of the transmitter reservoir storing information molecules \cite{khaloopour2019adaptive}. More importantly, scaling of the received concentration signals can also be observed in mobile MC systems \cite{ahmadzadeh2018stochastic, araz2023ratio}. For example, when the transmitter-receiver distance $d$ becomes smaller than the initial distance $d^\ast$, the received concentrations corresponding to both bit-$0$ and bit-$1$ increase, scaled by the same factor that depends on the distance, as determined by the CIR in \eqref{eq:MCimpulseresponse}. Conversely, when the transmitter and receiver move further apart after deployment, the received concentrations decrease. In such cases, the logarithm of the received concentrations shifts by the same amount with respect to the logarithm of the initially optimal dissociation constant $K_\mathrm{D,opt}^\ast$, such that the logarithmic distance between the received signals remains constant. 

\item \textbf{Shift of received concentration signals:} The received ligand concentrations corresponding to bit-$0$ and bit-$1$ can shift by an equal amount of ligand concentration. Examples scenarios where this may occur include high-rate MC channels where ISI substantially affects the received signals \cite{meng2014receiver}, or interference of background ligands that may arise from other MC links in the same channel (multi-user interference) or other biological processes producing the same type of ligand used for communication \cite{dinc2017theoretical, kuscu2022detection}. In such cases, the logarithms of the received concentrations shift by varying amounts with respect to the response curve and the logarithm of the initially optimal dissociation constant $K_\mathrm{D,opt}^\ast$. 
\end{itemize}

\begin{figure}[!t]
	\centering
	\includegraphics[width=0.90\columnwidth]{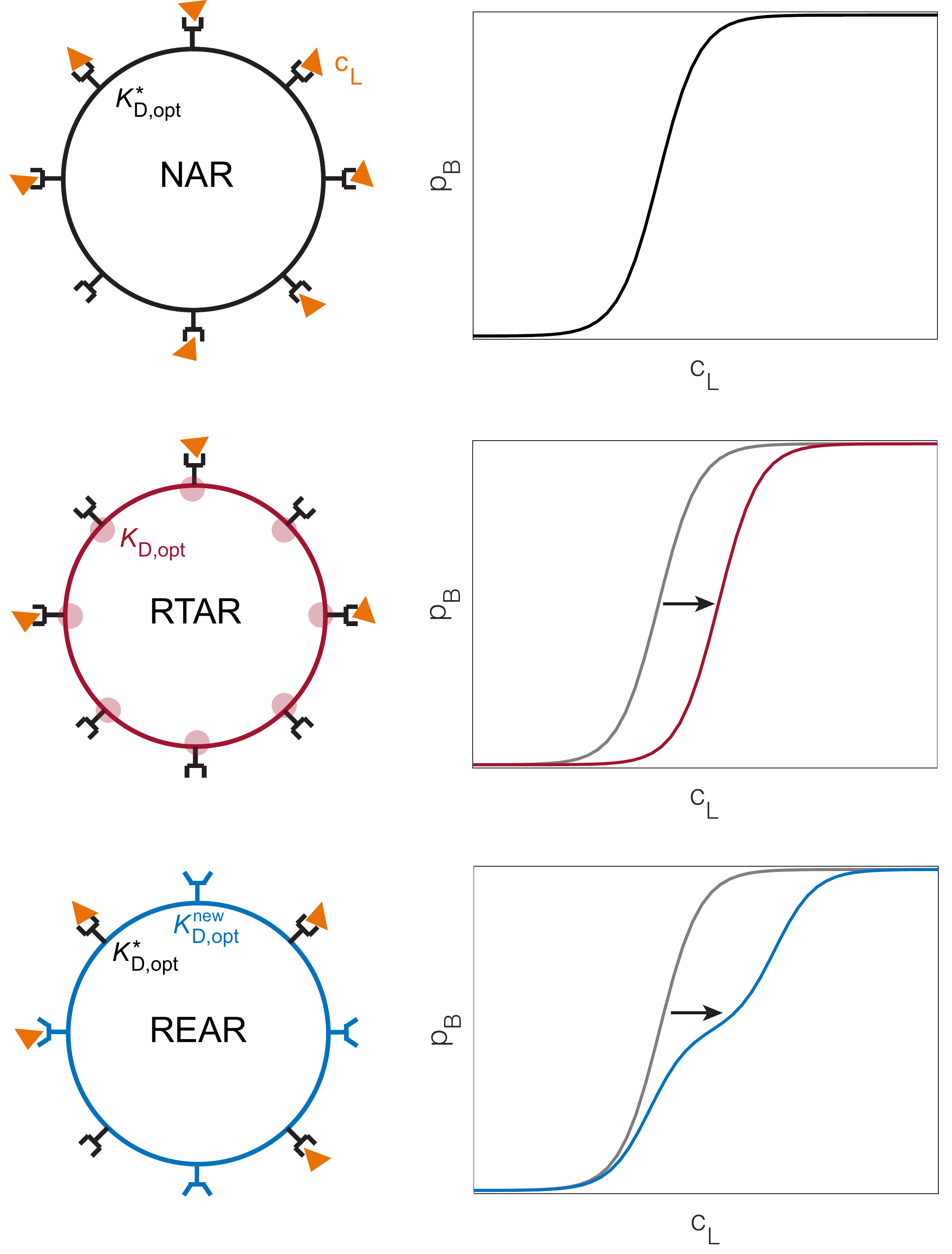}
	\caption{Conceptual demonstration of non-adaptive (NAR) and adaptive (RTAR \& REAR) MC receiver architectures together with their response curves.}
    \label{fig:RTARREAR}
\end{figure}

In this study, we consider three receiver architectures with ligand receptors (demonstrated in Fig. \ref{fig:RTARREAR}), two of which are adaptive: 
\begin{itemize}[leftmargin=*]
\item \textbf{Non-adaptive receiver (NAR):} Receiver with a response curve optimized for the baseline setting (with $K_\mathrm{D,opt}^\ast$). However, this type of receiver is unable to modify the response curve through its receptors to adapt to new conditions.  
\item \textbf{Receptor-tuning adaptive receiver (RTAR):} Receiver that can modify the response curve by tuning $K_\mathrm{D}$ of its membrane receptors, in order to maintain optimal detection performance under time-varying conditions. The corresponding optimal response curve is given by
\begin{align}
\p_\mathrm{B}(c_\mathrm{L}) = \frac{c_\mathrm{L}}{c_\mathrm{L} + K_\mathrm{D,opt}}.
\label{eq:pBRTAR}
\end{align}
\item \textbf{Receptor-expression adaptive receiver (REAR):}  
Receiver that can express a new type of receptors on its membrane, featuring a different dissociation constant, i.e., $K_\mathrm{D}^\mathrm{new}$, to extend its dynamic range and shift the  more sensitive region of the response curve to the new received concentration values attained by bit-$0$ and bit-$1$. In REAR, the response curve becomes:
\begin{align}
\p_\mathrm{B,2r}(c_\mathrm{L}) = \alpha \frac{c_\mathrm{L}}{c_\mathrm{L} + K_\mathrm{D, opt}^\mathrm{new}} + (1-\alpha) \frac{c_\mathrm{L}}{c_\mathrm{L} + K_\mathrm{D, opt}^\ast}, 
\label{eq:pBREAR}
\end{align}
where $K_\mathrm{D,opt}^\mathrm{new}$ is the optimal dissociation constant for the new receptors, and $\alpha$ is the number ratio of new receptors to all receptors. For simplicity, we assume $\alpha = 1/2$ for the remainder of this study. To avoid complications arising from an increase in the number of receptors—which would naturally lead to improved detection performance due to a higher number of independent samples \cite{kopuzlu2022capacity}—we assume that the total number of receptors remains constant, as if half of the original receptors were tuned to achieve $K_\mathrm{D,opt}^\mathrm{new}$. The derivation of $K_\mathrm{D,opt}^\mathrm{new}$ that maximizes error performance is, however, not analytically tractable; in this study, we obtain it by numerically solving the following optimization problem: 
\begin{align}
K_\mathrm{D,opt}^\mathrm{new} = \argmin_{K_\mathrm{D}^\mathrm{new}} \mathrm{BEP}.
\label{eq:KDoptimization}
\end{align}

\end{itemize}

In the following analyses, we assume all receiver types initially have receptors with optimal dissociation constants $K_\mathrm{D,opt}^\ast$. Time-varying conditions then lead to a loss of this optimality, which is addressed by adaptive receivers capable of modifying their response curves. The default values of system parameters used in the analyses are given in Table \ref{table:parameters}.  

\begin{table}[b!]
\centering
\renewcommand{\arraystretch}{1.5} 
\caption{Default values for system parameters}
\begin{tabular}{|p{6cm}|l|}
\hline
\textbf{Parameter} & \multicolumn{1}{l|}{\textbf{Default value}} \\
\hline
Number of receptors ($N_\mathrm{R}$) & 1000 \\
\hline
Number of molecules for bit-$1$ transmission ($N_{\mathrm{L}|1}$) & $5 \times 10^8$\\
\hline
Number of molecules for bit-$0$ transmission ($N_{\mathrm{L}|0}$) & $N_{\mathrm{L}|1}/50$\\
\hline
Diffusion coefficient of ligands ($D$) & $100 \, \mu \mathrm{m}^2/\mathrm{s}$\\
\hline
Transmitter-receiver distance ($d$) & $50 \, \mu \mathrm{m}$\\
\hline
Sampling time ($t_\mathrm{S}$) & $t_{\mathrm{Peak}}$\\
\hline
Probability of bit-$1$ transmission ($\p_1$) & $1/2$\\
\hline
Interference concentration ($c_\mathrm{int}$) & 0\\
\hline
\end{tabular}
\label{table:parameters}
\end{table}

\subsection{Scaling of Received Concentration Signals}
\label{sec:scaling}
First, we consider the scenario where the received concentration signal is scaled by a scaling constant $\gamma$ for both bit-$1$ and bit-$0$. Therefore, the scaled received signal can be represented in terms of the expected signal $c_{\mathrm{L}|s}^\ast$ as follows:
\begin{align}
c_{\mathrm{L}|s} = \gamma ~c_{\mathrm{L}|s}^\ast.
\label{eq:receivedsignal_scaled}
\end{align}
In this case, the optimal dissociation constant becomes
\begin{align}
K_\mathrm{D,opt} &= \sqrt{\gamma ~c_\mathrm{L|0}^\ast ~\gamma ~ c_\mathrm{L|1}^\ast } \\ \nonumber
&= \gamma ~K_\mathrm{D,opt}^\ast.
\label{eq:KDopt_scaling}
\end{align}

In the RTAR architecture, the receptors on the receiver surface are assumed to be capable of attaining $K_\mathrm{D,opt}$. In the REAR architecture, however, the receiver expresses a new type of receptors with dissociation constant $K_\mathrm{D,opt}^\mathrm{new}$, which is numerically obtained through the optimization process in \eqref{eq:KDoptimization}.

\begin{figure*}[t]
  \centering
  \subfigure[]{\label{fig:LR_scaling_001}\includegraphics[width=0.38\linewidth]{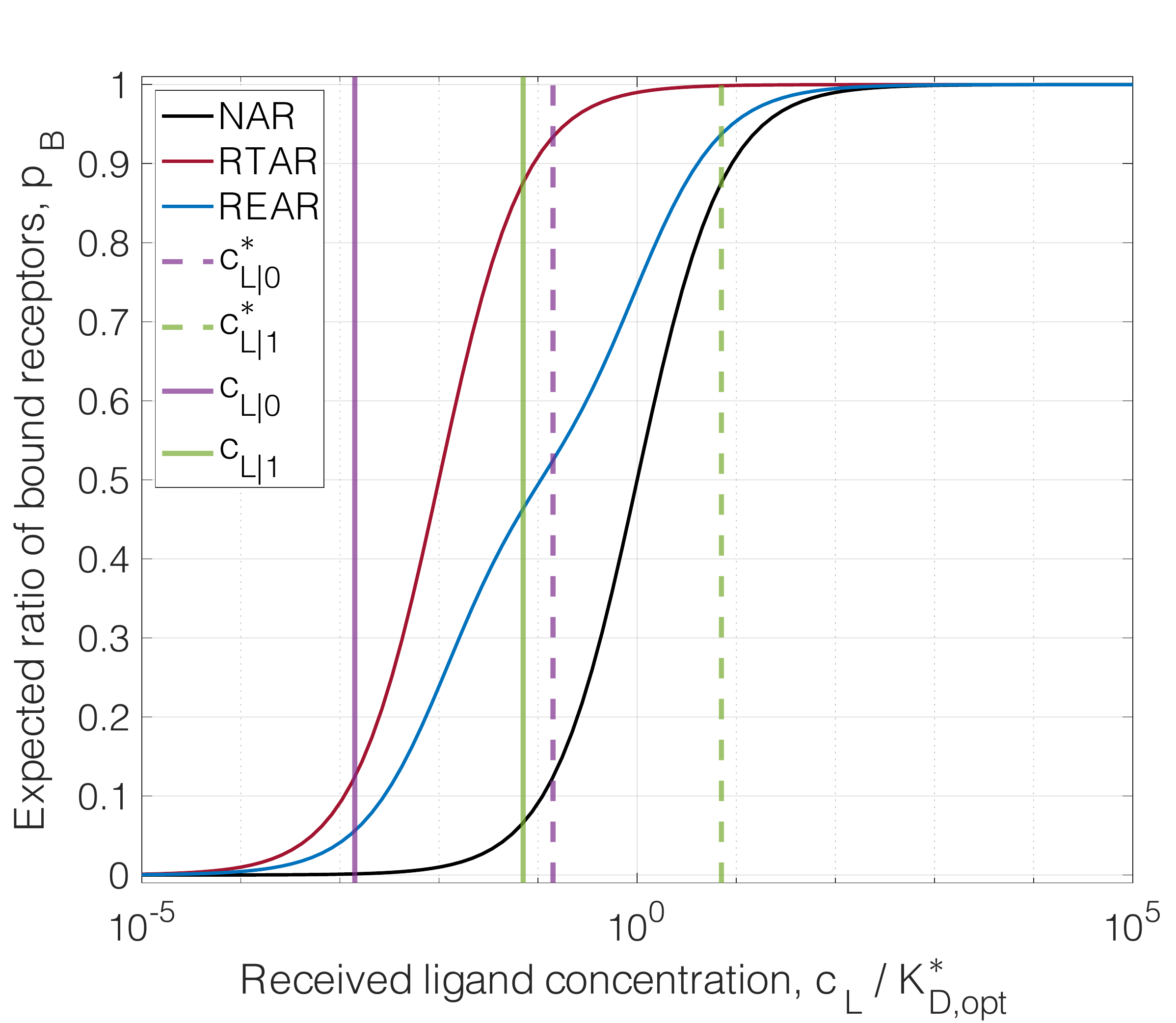}}   
  \subfigure[]{\label{fig:LR_scaling_100}\includegraphics[width=0.38\linewidth]{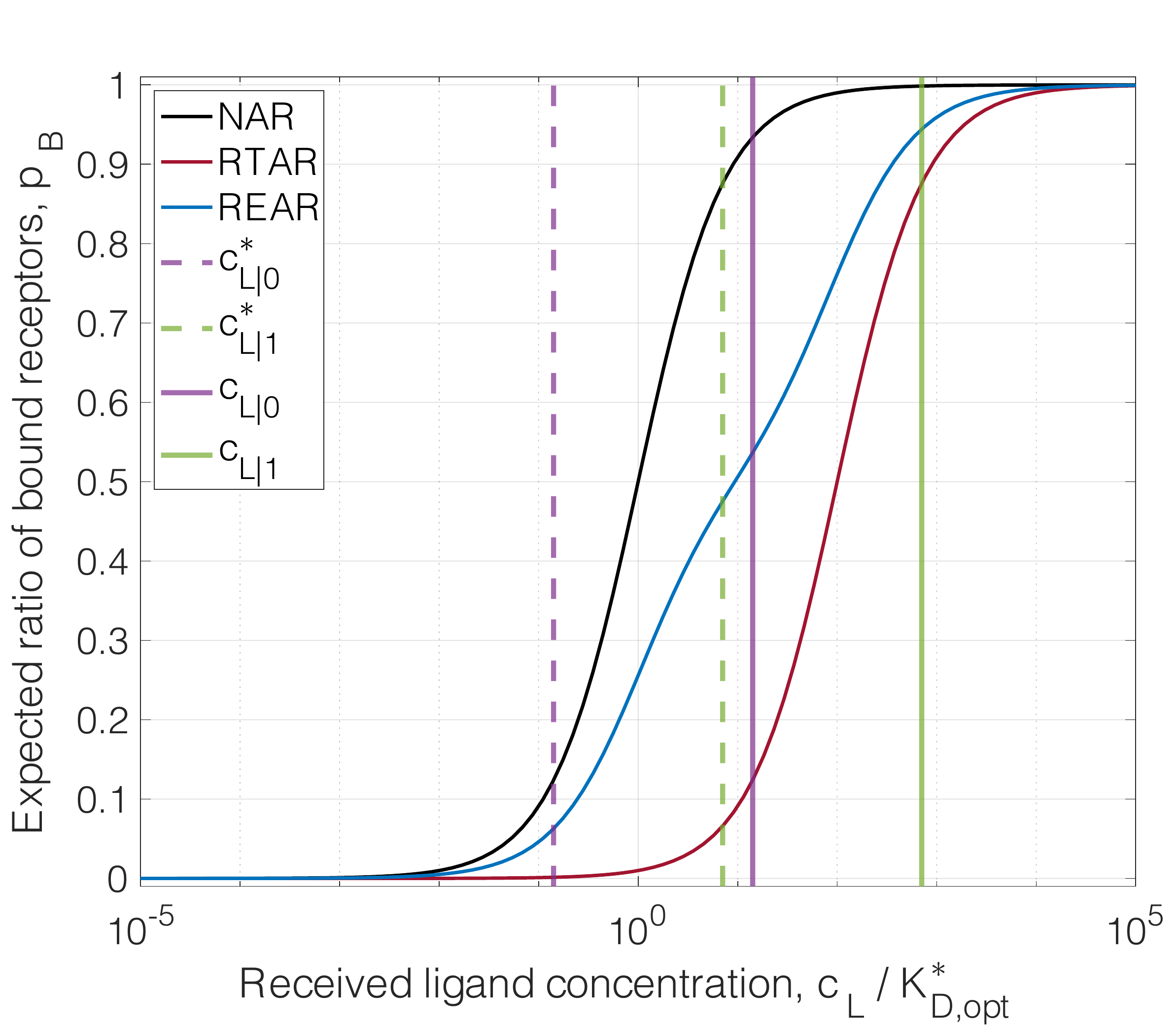}} 
  \subfigure[]{\label{fig:KD_scaling}\includegraphics[width=0.38\linewidth]{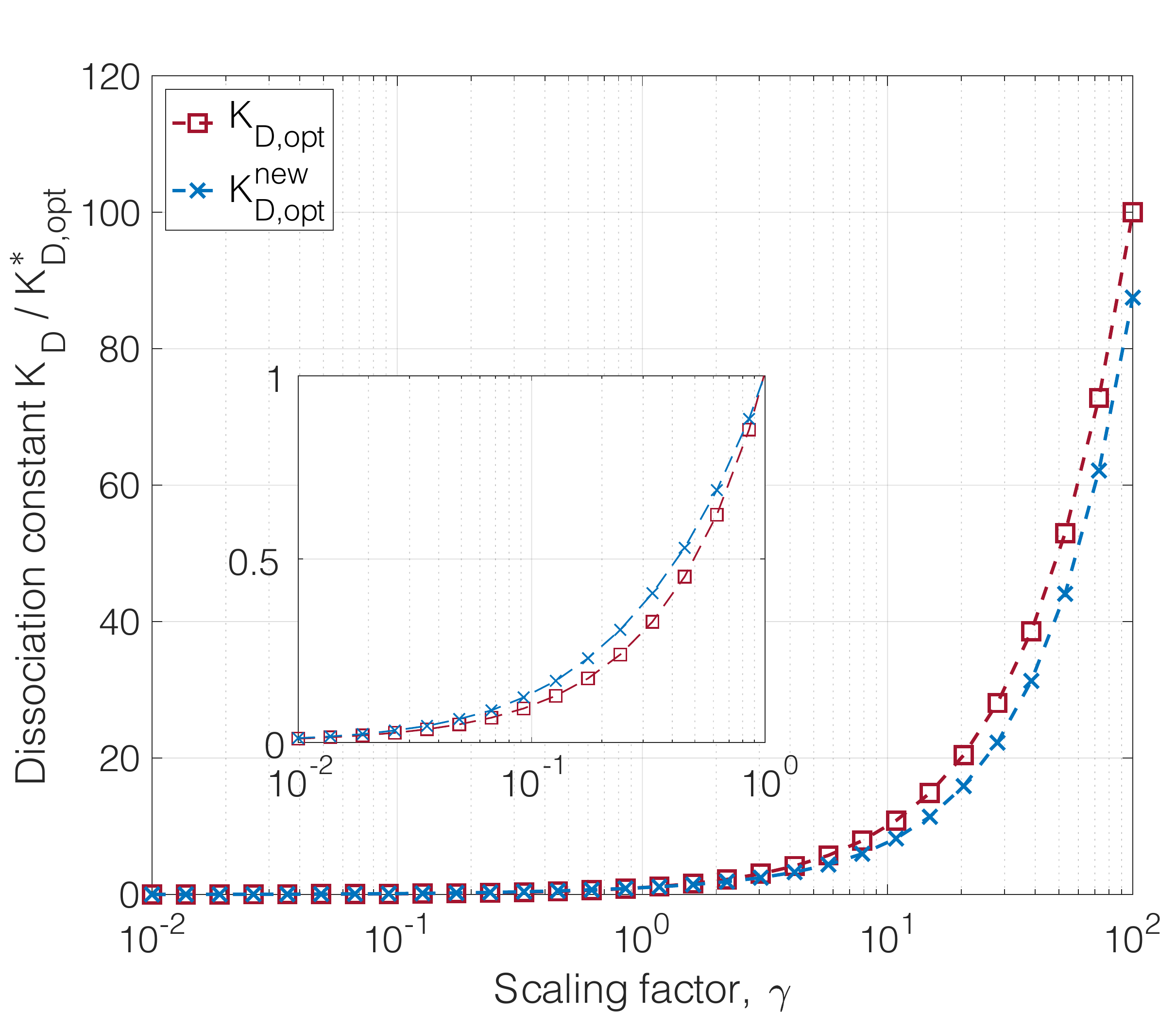}}   
  \subfigure[]{\label{fig:BEP_scaling}\includegraphics[width=0.38\linewidth]{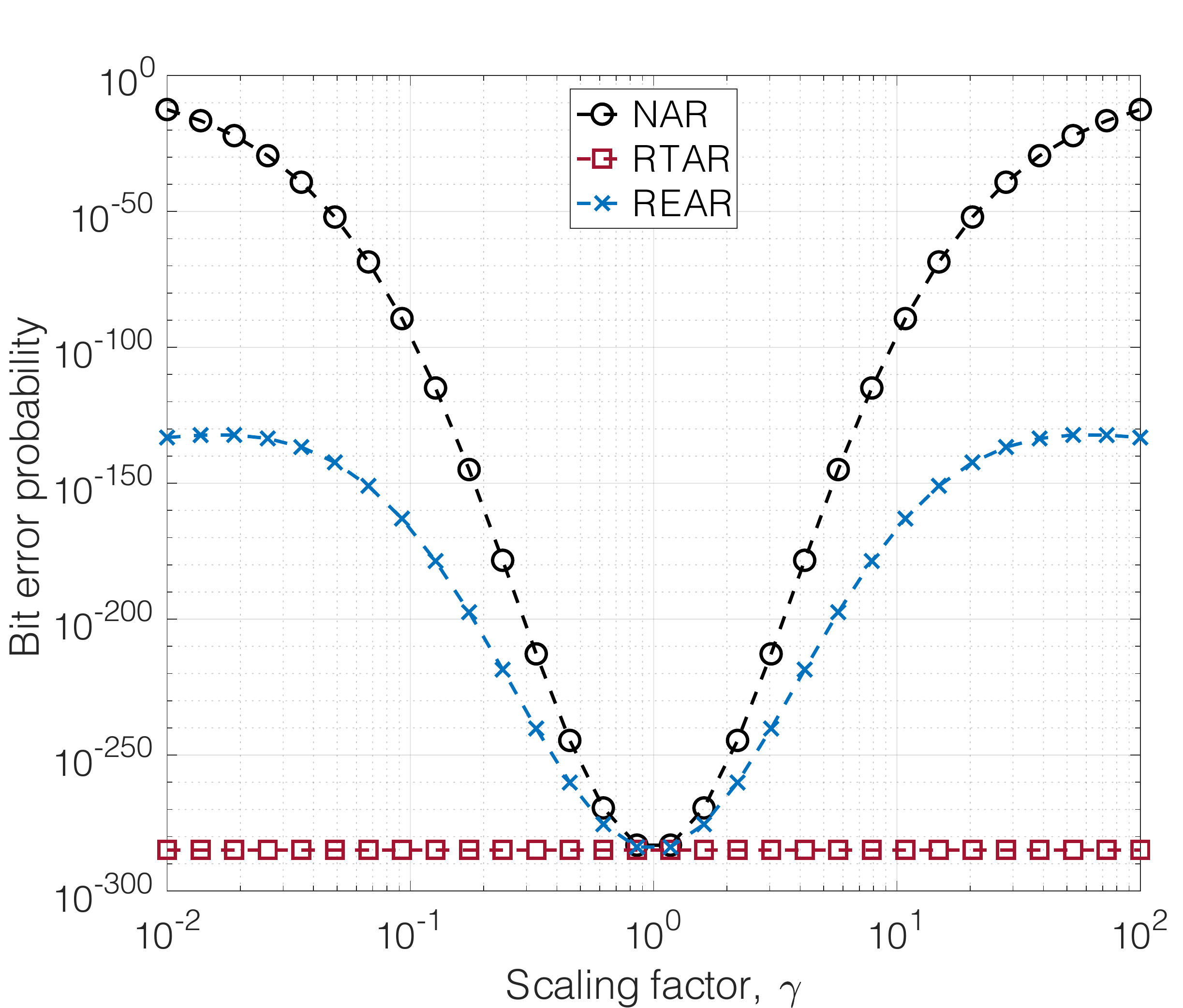}}  
  \caption{\emph{Adaptivity for scaling of received concentration signals:} (a, b) LR response curves for non-adaptive and adaptive receiver architectures when the scaling factor is $\gamma = 0.01$ and $\gamma = 100$, respectively. (c) Optimal dissociation constant $K_\mathrm{D,opt}$ of receptors for RTAR, and optimal dissociation constant of the newly expressed receptors $K_\mathrm{D,opt}^\mathrm{new}$ for REAR as a function of $\gamma$. (d) Corresponding BEP as a function of $\gamma$.}
\end{figure*}

The modified response curves, corresponding to these receptor adjustments, are presented in Figs. \ref{fig:LR_scaling_001} and \ref{fig:LR_scaling_100}, for $\gamma = 0.01$ and $\gamma = 100$, respectively. The vertical dashed lines indicate the initial received concentrations corresponding to bit-$0$ and bit-$1$ in the baseline setting, centered around $c_\mathrm{L} = K_\mathrm{D,opt}^\ast$, while the vertical continuous lines represent the scaled received concentrations in the new setting. When $\gamma = 0.01$, the scaled concentrations are positioned to the left of the baseline concentrations, while for $\gamma = 100$, they are placed to their right. As observed, RTAR effectively tunes the response curve to center itself with respect to the scaled received concentrations in logarithmic scale. In contrast, REAR exhibits limited adaptability in adjusting the response curve, as the presence of original non-tunable receptors compromises the detection performance. 

The corresponding BEP analysis, employing the optimal $K_\mathrm{D,opt}$ and $K_\mathrm{D,opt}^\mathrm{new}$, for RTAR and REAR, respectively, is provided in Fig. \ref{fig:BEP_scaling} as a function of the scaling factor $\gamma$. Significant adaptivity gains are evident in both RTAR and REAR cases. In particular, RTAR's error performance remains unaffected by the scaling factor, highlighting the potential of this adaptive architecture for mobile MC systems and scenarios involving degrading enzymes within the MC channel.

\begin{figure*}[t]
  \centering
  \subfigure[]{\label{fig:LR_shift_2}\includegraphics[width=0.38\linewidth]{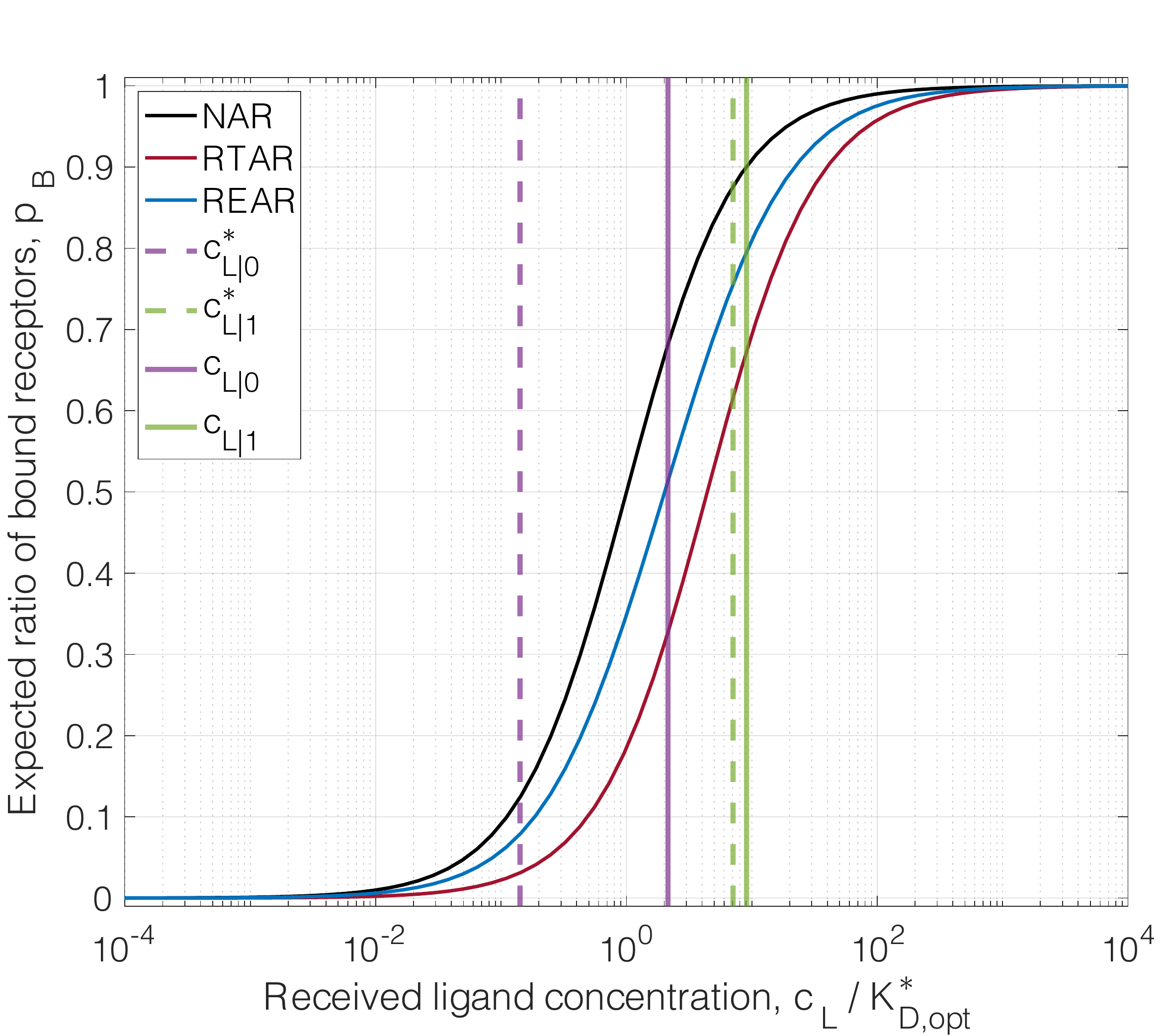}} 
  \subfigure[]{\label{fig:LR_shift_20}\includegraphics[width=0.38\linewidth]{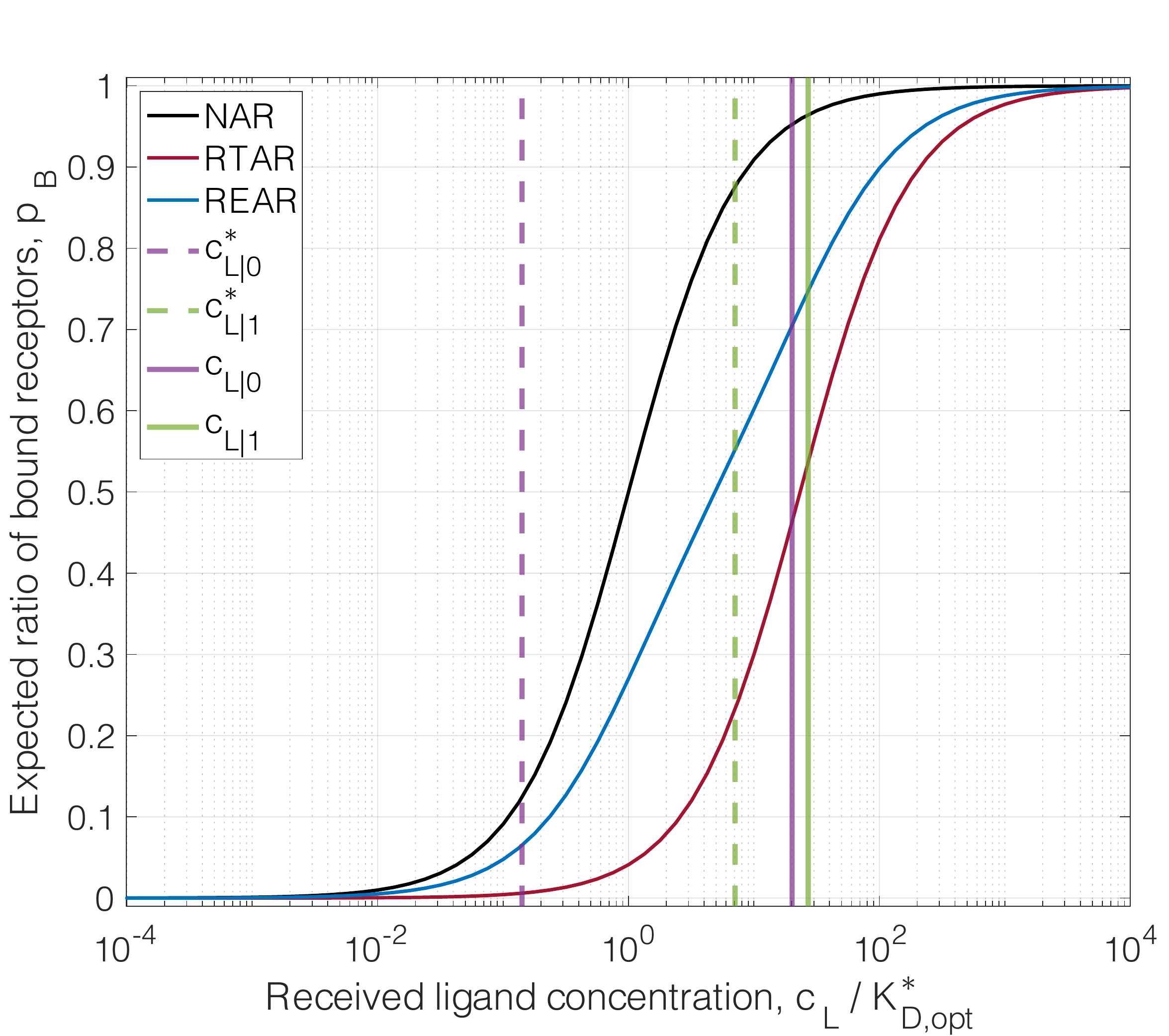}} 
  \subfigure[]{\label{fig:KD_shift}\includegraphics[width=0.38\linewidth]{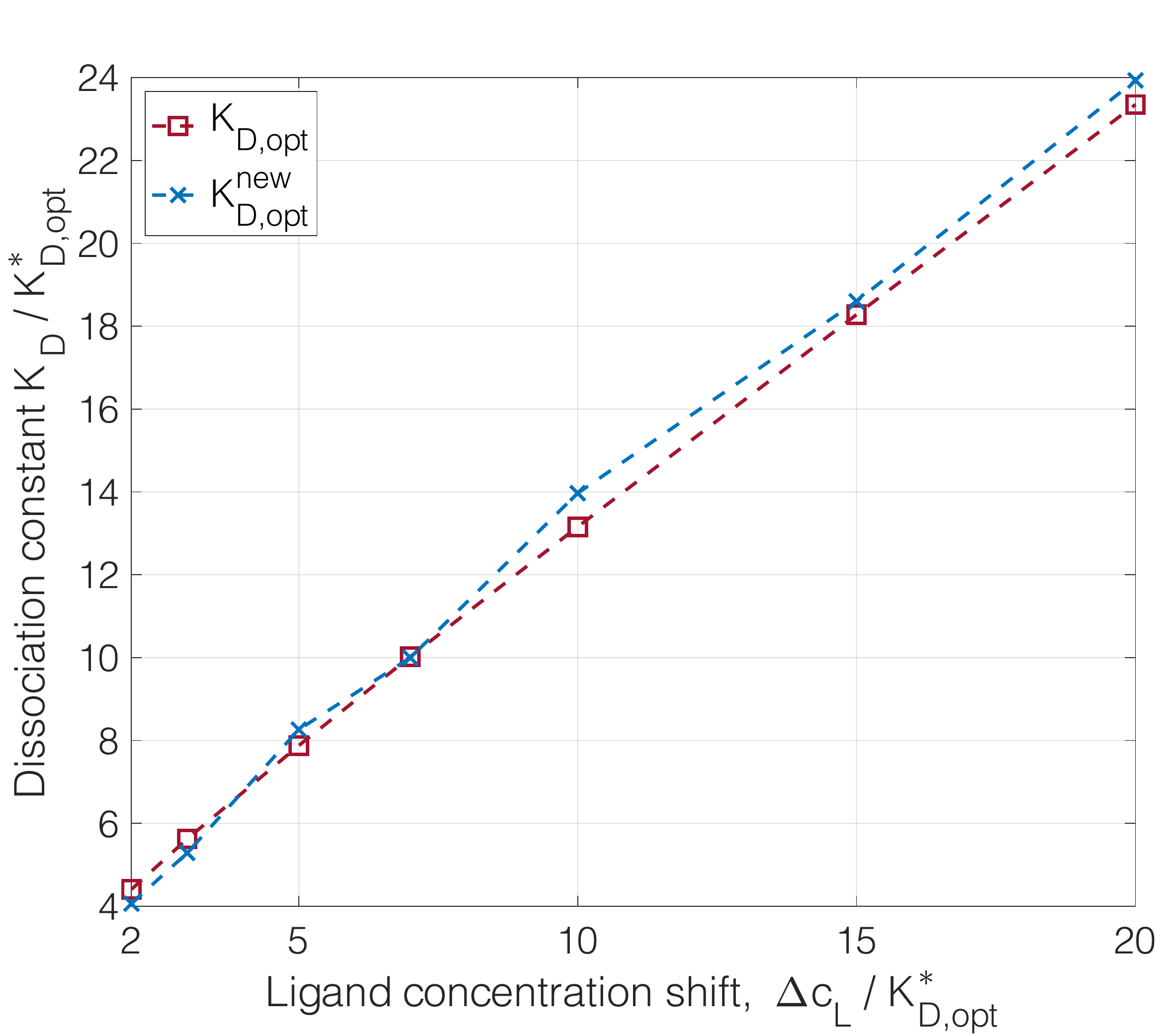}} 
  \subfigure[]{\label{fig:BEP_shift}\includegraphics[width=0.38\linewidth]{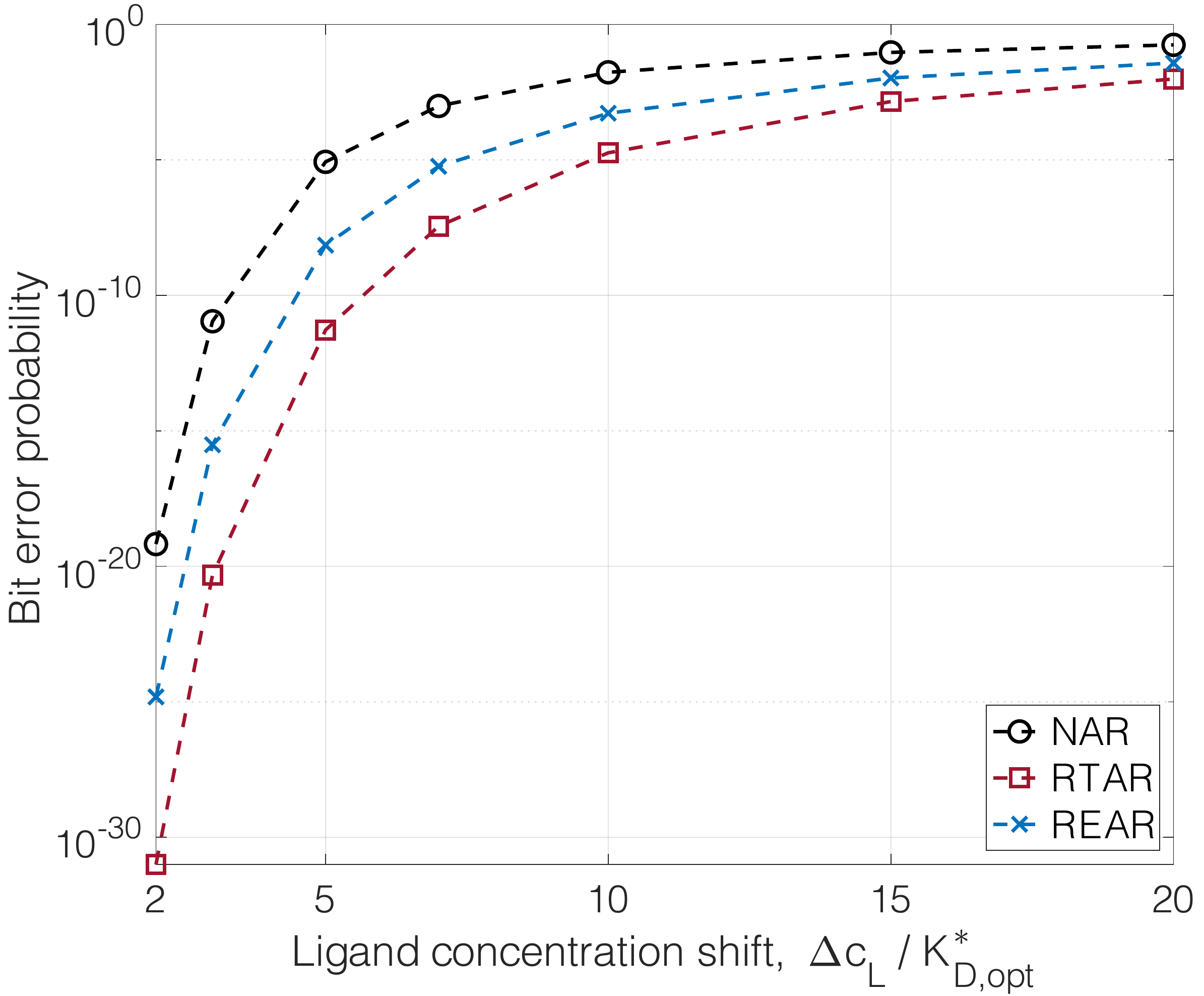}} 
  \caption{\emph{Adaptivity for shift of received concentration signals:} (a, b) LR response curves for non-adaptive and adaptive receiver architectures when the concentration shift is $\Delta c_\mathrm{L} = 2 \times K_\mathrm{D,opt}^\ast$ and $\Delta c_\mathrm{L} = 20 \times K_\mathrm{D,opt}^\ast$, respectively. (c) Optimal dissociation constant $K_\mathrm{D,opt}$ of receptors for RTAR, and optimal dissociation constant of the newly expressed receptors $K_\mathrm{D,opt}^\mathrm{new}$ for REAR as a function of $\Delta c_\mathrm{L}$. (d) Corresponding BEP as a function of $\Delta c_\mathrm{L}$.}
\end{figure*}

\subsection{Shift of Received Concentration Signals} 
\label{sec:shift}

In this set of scenarios, we examine the shift of received concentration signals by the same amount of ligand concentration $\Delta c_\mathrm{L}$, resulting in the following received signal form:
\begin{equation}
c_\mathrm{L|s} = c_\mathrm{L|s}^\ast + \Delta c_\mathrm{L} ~~~\text{with} ~ \Delta c_\mathrm{L} > 0. 
\label{eq:concentration_shift}
\end{equation}
The optimal dissociation constant is then given by 
\begin{align}
K_\mathrm{D,opt} = \sqrt{(c_{\mathrm{L}|0}^\ast + \Delta c_\mathrm{L} )(c_{\mathrm{L}|1}^\ast + \Delta c_\mathrm{L} )}.
\label{eq:KDopt_shift}
\end{align}

The baseline and shifted received concentrations and the corresponding modified response curves for the RTAR and REAR architectures are shown in Figs. \ref{fig:LR_shift_2} and \ref{fig:LR_shift_20}, for $\Delta c_\mathrm{L} = 2 \times K_\mathrm{D,opt}^\ast$ and $\Delta c_\mathrm{L} = 20 \times K_\mathrm{D,opt}^\ast$, respectively. The figures demonstrate that concentration shifts cause the received concentrations for bit-$0$ and bit-$1$ to become closer to each other on logarithmic scale. When $\Delta c_\mathrm{L} = 20 \times K_\mathrm{D,opt}^\ast$, received concentrations for both bit-$0$ and bit-$1$ fall within the saturation region of the receiver (i.e., $\p_\mathrm{B} > 0.9$), indicating an increased difficulty for the receiver to distinguish between them. The RTAR architecture successfully circumvents the effects of the shift by tuning the dissociation constant of the receptors to its optimal value, $K_\mathrm{D,opt}$. In contrast, the REAR architecture expresses a new type of receptor to extend the dynamic range towards higher concentration values, albeit limited by the presence of the original non-tunable receptors. The optimal dissociation constants attained by the tunable receptors in RTAR and the newly expressed receptors in REAR are illustrated in Fig. \ref{fig:KD_shift} for varying concentration shifts. For both architectures, we observe a nearly linear increase in the optimal dissociation constants as the shift increases. 

\begin{figure}[b!]
 	\centering
 	\includegraphics[width=0.8\columnwidth]{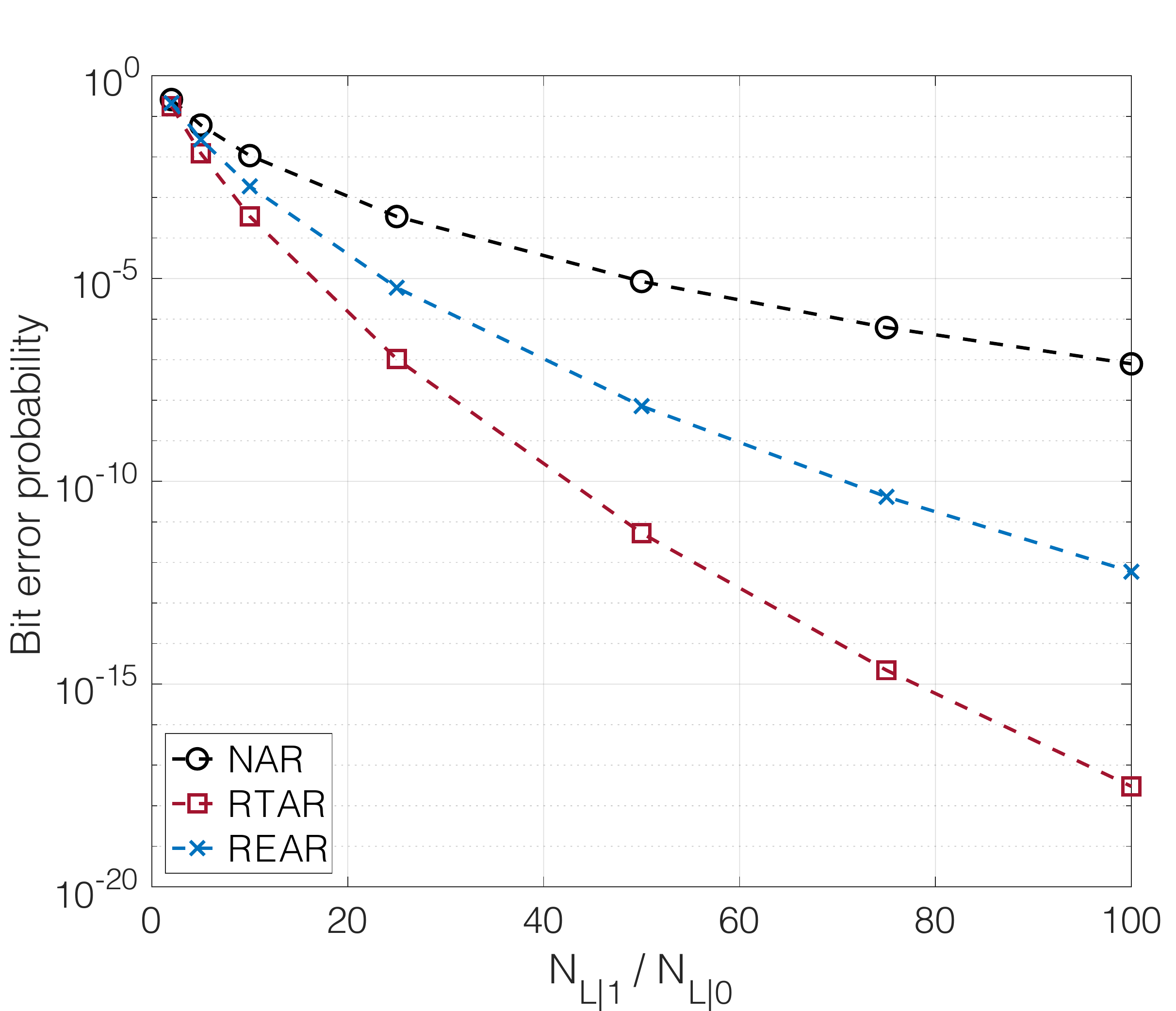}
 	\caption{Impact of the ratio between the transmitted number of molecules for bit-$1$ and bit-$0$, $N_\mathrm{L|1} / N_\mathrm{L|0}$, on the error performance.}
 	\label{fig:shift_BEP_varying_ratio}
 \end{figure}

Performance analysis in terms of BEP is presented in Fig. \ref{fig:BEP_shift}. In this analysis, we observe that RTAR achieves the best performance, while the improvement observed in REAR is still significant compared to the NAR architecture. Additionally, we evaluate the impact of the ratio between the transmitted number of molecules for bit-$1$ and bit-$0$, which corresponds to the ratio of received concentrations for bit-$1$ and bit-$0$ in the baseline setting. As indicated in Table \ref{table:parameters}, the default ratio is set to $50$. As observed in Fig. \ref{fig:shift_BEP_varying_ratio}, this ratio has a considerable impact on the error performance of all receiver architectures, including NAR. However, the relative improvement introduced by the adaptive architectures in comparison to NAR remains largely unchanged, suggesting that the concentration ratio plays an insignificant role in the adaptivity analysis. The performance of RTAR and REAR in the presence of shifts in the received concentration signals will be further assessed under practical scenarios, such as ISI and stochastic background interference, in the following section. 
 
\begin{figure*}[t!]
  \centering
\subfigure[]{\label{fig:BEP_enzymes}\includegraphics[width=0.32\linewidth]{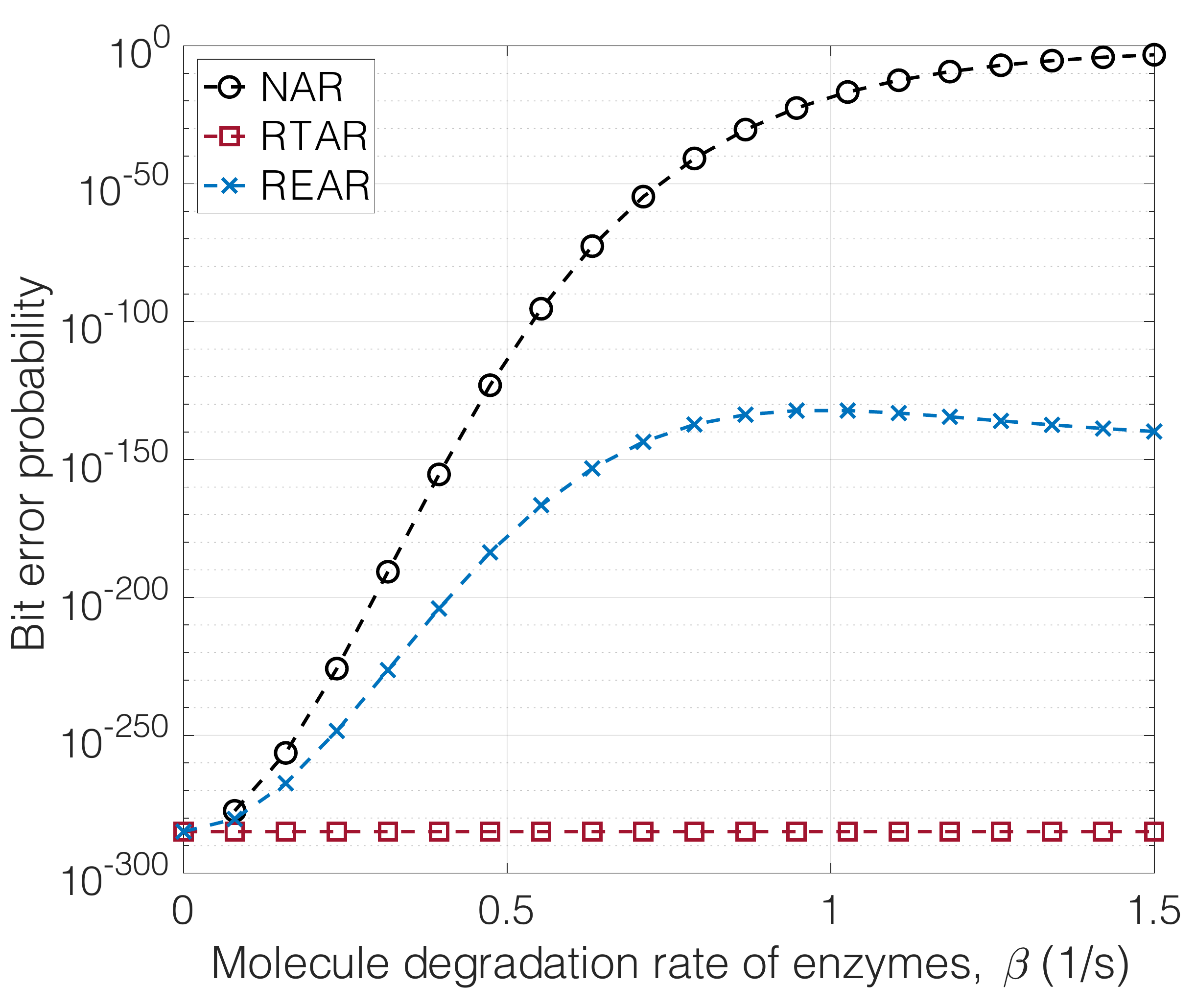}} 
  \subfigure[]{\label{fig:LR_enzymes_0.71}\includegraphics[width=0.32\linewidth]{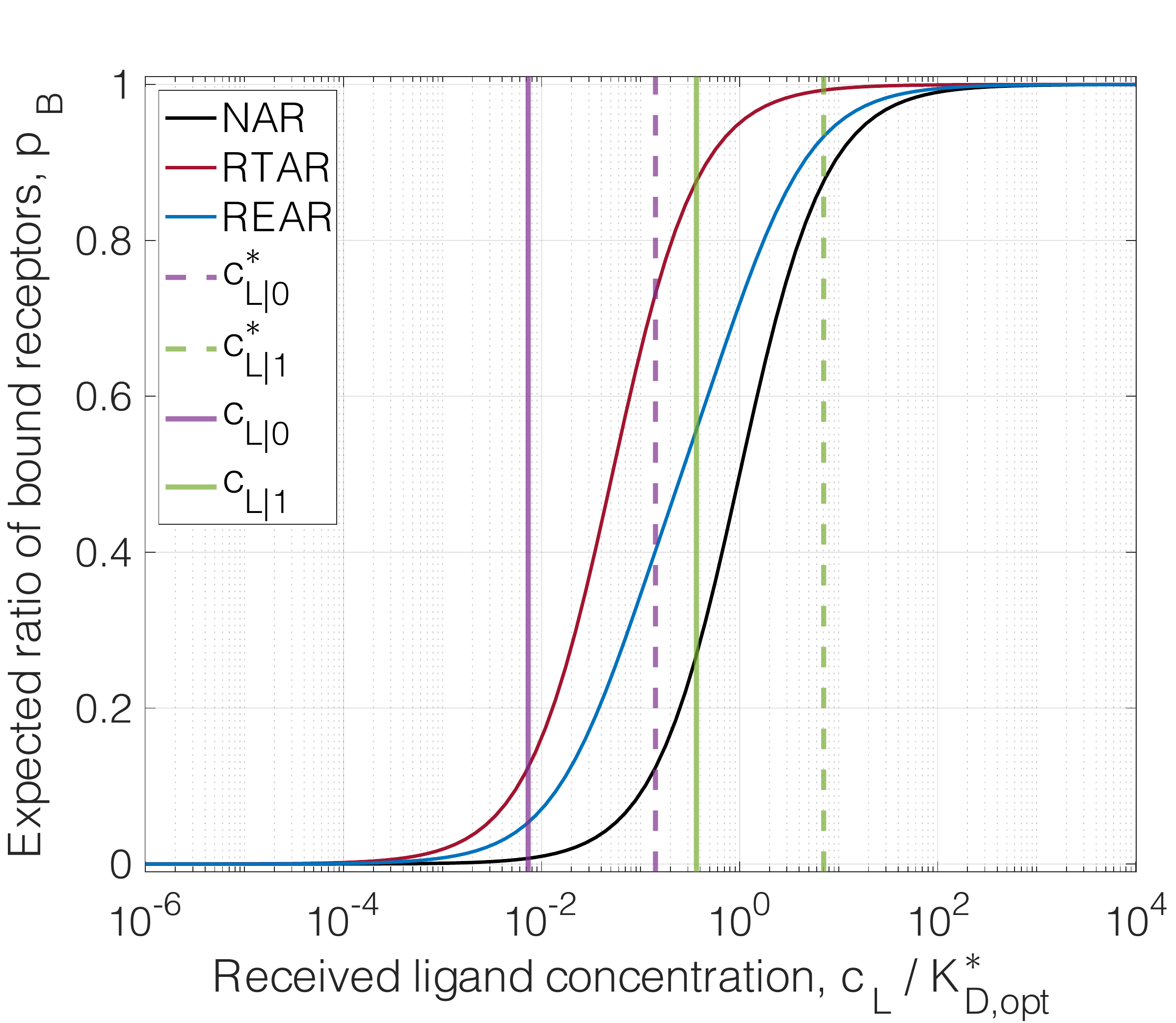} } 
\subfigure[]{\label{fig:LR_enzymes_1.5}\includegraphics[width=0.32\linewidth]{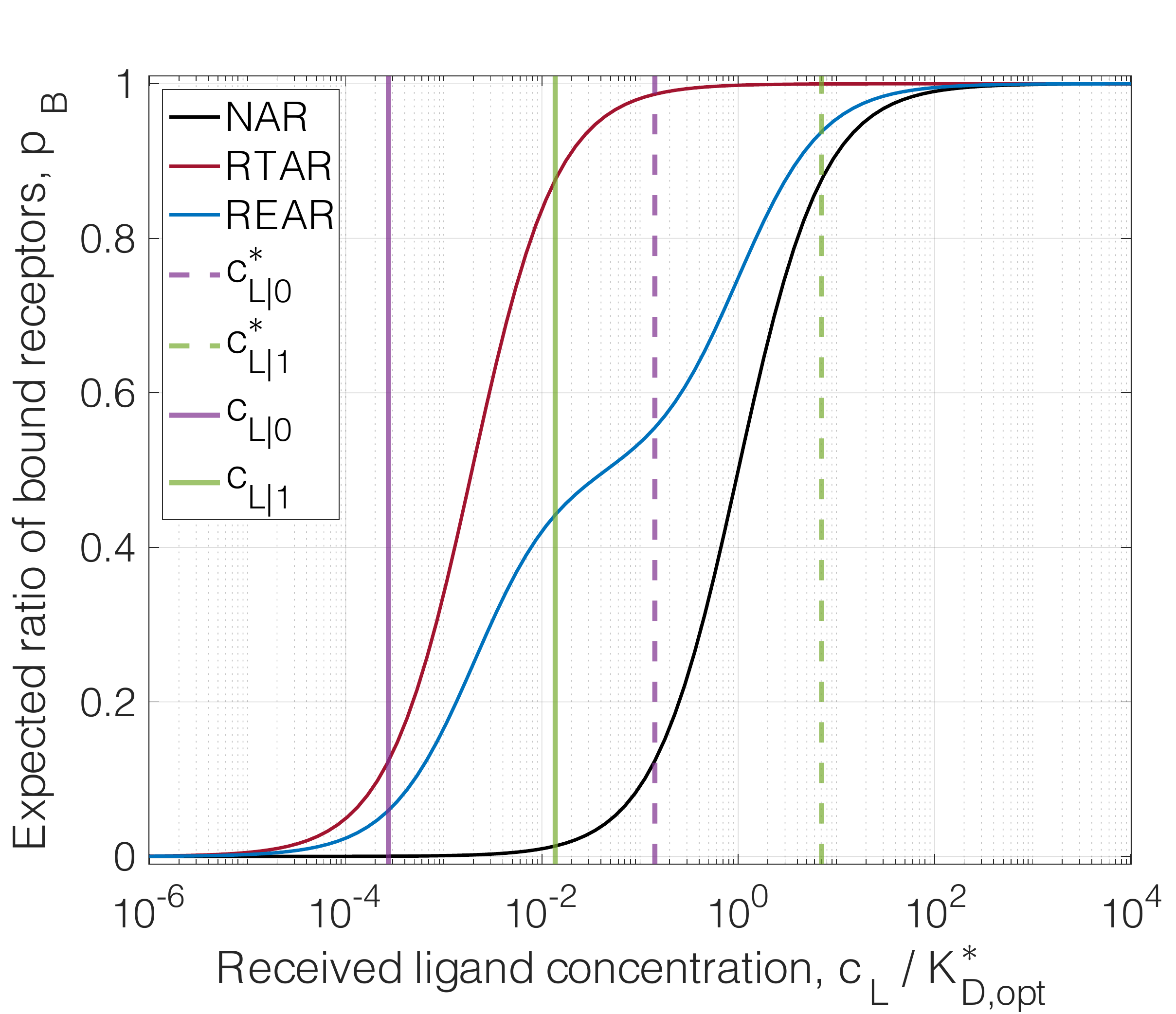}} 
  \caption{\emph{Adaptivity for enzymatic degradation of ligands:} (a) BEP as a function of $\beta$. (b, c) LR response curves for non-adaptive and adaptive receiver architectures when the degradation rate is $\beta = 0.71 \mathrm{s}^{-1}$ and $\beta = 1.5 \mathrm{s}^{-1}$, respectively.}
\end{figure*}

\section{Adaptive Receivers in Practical MC Scenarios}
\label{sec:practical}
The evaluation of the adaptive receiver architectures in the previous section provided insights into the performance of RTAR and REAR in general cases of scaling and shift of the received concentration signals. In this section, we examine these architectures in the context of practical time-varying MC scenarios that have been extensively studied in the literature, including ligand degradation by enzymes in the channel, ISI, and stochastic background interference. 

\subsection{Case I: Enzymatic Degradation of Ligands}
\label{sec:enzymes}

Biological environments, where most MC applications are expected to be deployed, are typically characterized by an abundance of various types of molecules, including proteins and enzymes. Certain molecules, like enzymes, can react with information-carrying ligands, changing their chemical structure and rendering them unrecognizable by the receiver. This reaction effectively removes the ligands from the communication system. The impact of degrading enzymes has been widely explored in the MC literature, and studies have demonstrated that their presence can improve MC performance by reducing channel memory and, consequently, ISI \cite{noel2014improving}. Therefore, researchers have proposed intentionally deploying enzymes near transmitters or receivers to improve performance \cite{cho2017effective}. The presence of degrading enzymes, however, alters the received concentration signals. 

In this analysis, we consider a memoryless channel and ignore ISI. We assume that a single type of enzyme can degrade the transmitted ligands in the channel. The impact of enzymes on the ligand concentration can be formulated by updating CIR as follows \cite{cho2017effective}:  
\begin{align}
h(t,d) =  h^\ast(t,d) ~e^{- \beta t},
\label{eq:cnew_enzyme}
\end{align}
where $\beta$ denotes the enzyme degradation rate. Accordingly, the modified received signals at the sampling time corresponding to bit-$0$ and bit-$1$ can be represented as 
\begin{align}
c_{\mathrm{L}|s} = c_{\mathrm{L}|s}^\ast ~ e^{- \beta t_\mathrm{S}}.
\label{eq:cnew_enzyme}
\end{align}
Here, we can notice that the scaling factor discussed in Section \ref{sec:scaling} is now given by $\gamma = e^{- \beta t_\mathrm{S}} $. Therefore, we can expect the performance of adaptive receivers in terms of BEP to be comparable to the analysis of received signal scaling presented in Section \ref{sec:scaling}. 

The performance of adaptive receiver architectures in the presence of enzymes with varying degradation rates is evaluated in comparison to the non-adaptive receiver. The results are presented in Fig. \ref{fig:BEP_enzymes}. As expected, the RTAR architecture demonstrates the ability to keep pace with the scaling of the received concentrations and adapts its response curve by optimally tuning the dissociation constant $K_D$ of its receptors, which is evidenced by Figs. \ref{fig:LR_enzymes_0.71} and \ref{fig:LR_enzymes_1.5}. In contrast, the performance of NAR architecture rapidly declines as enzyme degradation rate increases. 

The performance trajectory of REAR presents a distinct pattern. As shown in Fig. \ref{fig:BEP_enzymes}, it initially deteriorates in response to an increasing degradation rate, while still offering considerable improvement compared to NAR. This trend subsequently reverses around $\beta \sim 0.75-1.00 \mathrm{s}^{-1}$, where REAR performance experiences a slight increase and ultimately plateaus as the degradation rate approaches $\beta \sim 1.5 \mathrm{s}^{-1}$. This atypical path can be better understood by examining the evolution of the response curve through a comparison of Fig. \ref{fig:LR_enzymes_0.71} ($\beta = 0.71 \mathrm{s}^{-1}$) and Fig. \ref{fig:LR_enzymes_1.5} ($\beta = 1.5 \mathrm{s}^{-1}$). At low and moderate degradation rates, the influence of original receptors with non-optimal dissociation constants remains pronounced on the combined response curve, which is visible particularly at low occupation ratios (low $\p_B$). However, as the degradation rate increases, the received concentrations shift further toward lower values, diminishing the impact of original receptors. As a result, the dissociation constant of newly expressed receptors, $K_\mathrm{D,opt}^\mathrm{new}$, can independently track the scaled concentration levels, leading to a slight improvement in performance. 

\subsection{Case II: Intersymbol Interference (ISI)}
\label{sec:isi}

MC channels are notorious for their substantial memory, which stems from the persistence of transmitted ligands from earlier transmissions within the channel. This results in ISI, which occurs when the ligands are not absorbed by the receiver or not degraded by enzymes in the channel \cite{jamali2019channel}. ISI increases as the signaling interval $T_\mathrm{S}$ decreases, thereby limiting the data rate. Typically, when $T_\mathrm{S}$ is sufficiently large, ISI is neglected to simplify the analyses \cite{kuscu2019transmitter}. 

\begin{figure*}[t]
  \centering
  \subfigure[]{\label{fig:BEP_isi_Ts}\includegraphics[width=0.37\linewidth]{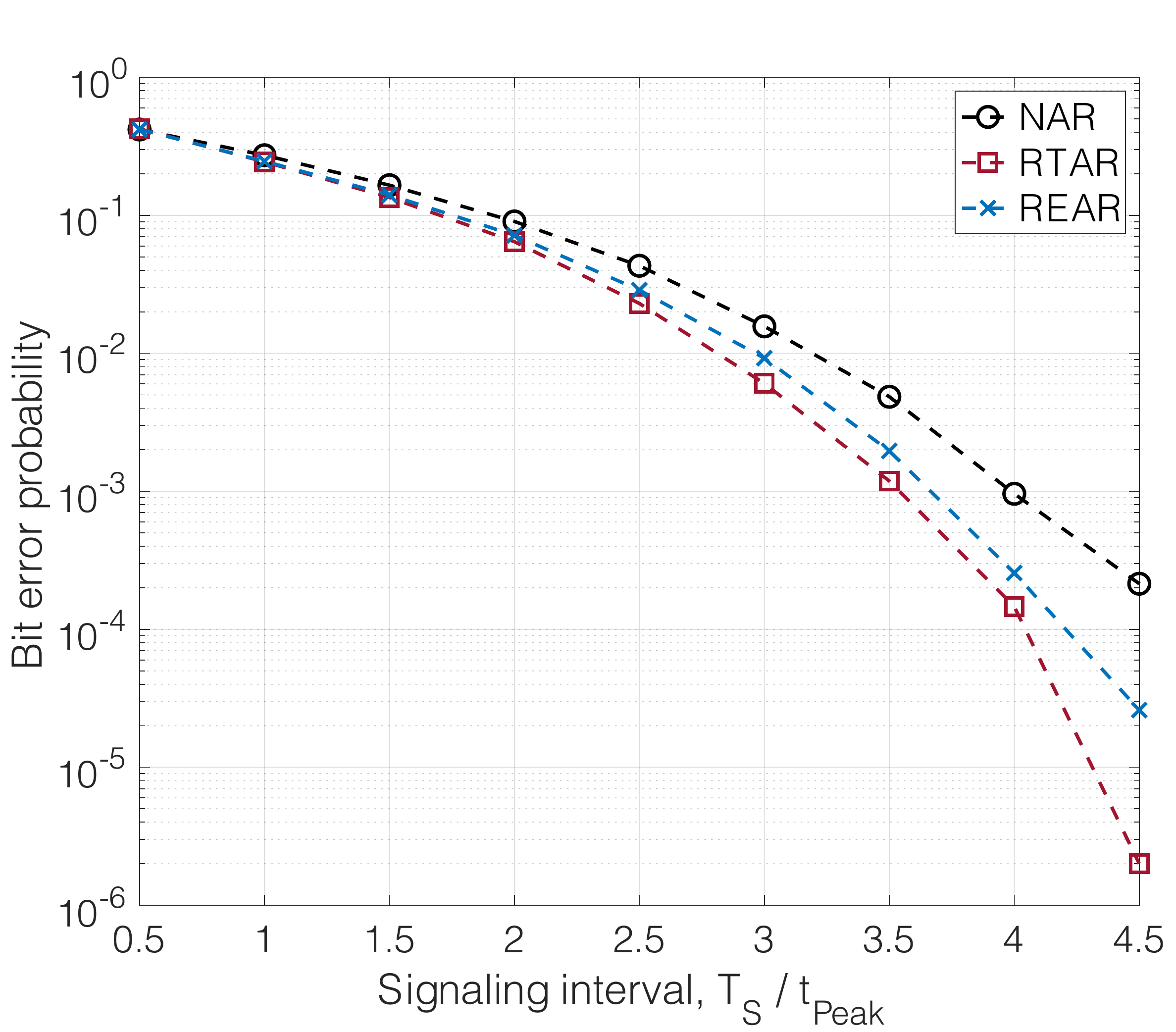}}
  \subfigure[]
  {\label{fig:BEP_isi_memory} \includegraphics[width=0.37\linewidth]{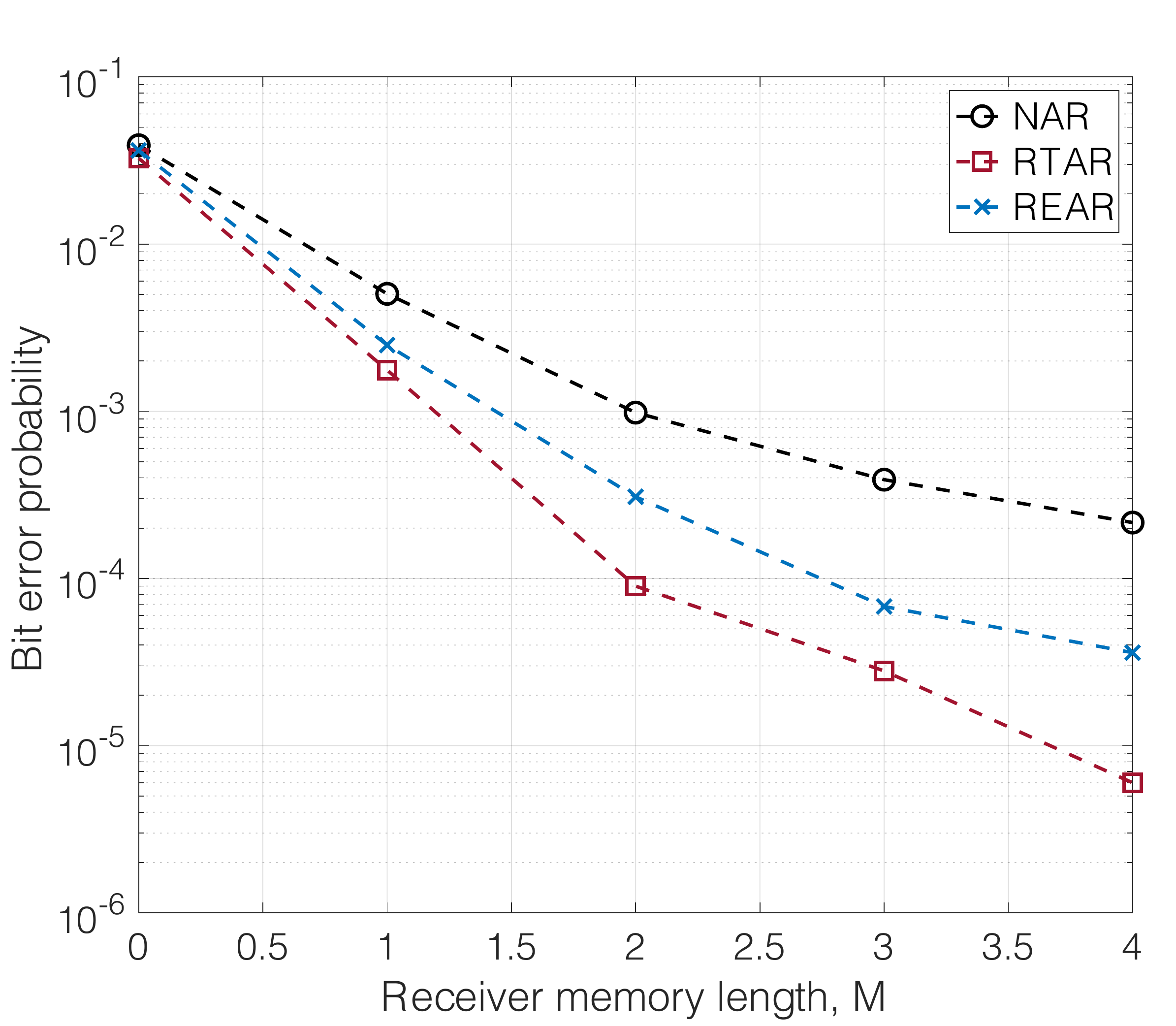}} 
  \caption{\emph{Adaptivity for ISI:} BEP as a function of (a) signaling interval, and (b) receiver memory length for non-adaptive and adaptive receiver architectures. }
\end{figure*}

To address the issue of ISI, researchers have developed various detection techniques that differ significantly in terms of complexity \cite{kuscu2019transmitter}. In this section, we demonstrate that practical adaptive receiver architectures can effectively help further reduce ISI and enhance data rates. We assume that the receiver is initially equipped with receptors featuring optimal dissociation constants for a baseline scenario with no ISI, and subsequently exposed to conditions where ISI becomes significant. This scenario may arise, for instance, when $T_\mathrm{S}$ is initially large, but later on, the system is required to increase data transmission rates by reducing $T_\mathrm{S}$, ultimately resulting in increased ISI. To investigate the impact of ISI, we employ a discrete-time model of the MC channel: 
\begin{align} \nonumber
c_{\mathrm{L}|s[i]} &= N_\mathrm{L|s[i]} h^\ast(t_\mathrm{S},d) + \sum_{j = i - I}^{i-1} N_{L|s[j]} h^\ast\left(t_\mathrm{S} + (i-j) T_S, d\right) \\ \nonumber
&= c_{\mathrm{L}|s[i]}^\ast + \Omega[i], 
\label{eq:received_signal_isi}
\end{align}
where the square-bracketed indices denote the time indices of transmitted symbols, with $i$ representing the index of the current symbol. The length of the channel memory, denoted as $I$, is assumed to be finite on the grounds that the impact of earlier transmissions—signals transmitted before ($i-I$)—diminishes over time and becomes negligible compared to most recent signal contributions \cite{meng2014receiver, kuscu2018maximum}. In this model, it is important to note that the actual received signal is composed of two parts: the expected received signal in the baseline setting, $c_{\mathrm{L}|s[i]}^\ast$, and an ISI term $\Omega[i]$. The presence of this ISI term disrupts the optimality of the receptors that were initially set for the baseline setting without ISI. 

In this analysis, we assume that the receiver has a memory with capacity of $M$ bits, which stores the $M$ most recently decoded bits to estimate the ISI value \cite{kuscu2018maximum}. The receiver's ISI estimate, denoted as $\hat{\Omega}_M[i]$, can be expressed as follows 
\begin{align}
\hat{\Omega}_M[i] &= \sum_{j=i-M}^{i-1} N_{L|\hat{s}[j]}  ~ h^\ast(t_\mathrm{S} + (i-j) T_S,d) \\ \nonumber
&~~~ + \left(\p_0 N_{L|0} + \p_1 N_{L|1}\right) \\ \nonumber
&~~~ \times \sum_{j=i-I}^{i-M-1} h^\ast(t_\mathrm{S} + (i-j) T_S,d).
\label{eq:isi_estimate}
\end{align}
In this memory-based estimation, the receiver calculates the expected value of ISI from previous symbols that are beyond the memory length, by using the transmission probabilities of bit-$0$ ($\p_0$) and bit-$1$ ($\p_1$). As a result, the estimated received concentration signal for each bit are given as 
\begin{equation}
\hat{c}_{\mathrm{L}|s} = c_{\mathrm{L}|s}^\ast + \hat{\Omega}_M[i]. 
\label{eq:received_signal_estimate}
\end{equation}

\begin{table}[b]
\centering
\renewcommand{\arraystretch}{1.5} 
\caption{Default parameter values for ISI analysis}
\begin{tabular}{|p{5cm}|l|}
\hline
\textbf{Parameter} & \multicolumn{1}{l|}{\textbf{Default value}} \\
\hline
Channel memory length  ($I$) & 30 \\
\hline
Receiver memory length ($M$) & $2$ bits\\
\hline
Length of signaling interval ($T_\mathrm{S}$) & $4 \times t_\mathrm{Peak}$\\
\hline
\end{tabular}
\label{table:isi_parameters}
\end{table}

All the receiver architectures, including NAR, dynamically adjust their binary decision thresholds (see Appendix \ref{AppendixA} for its derivation) based on the received concentration estimate, which is updated at each signaling interval. However, it would be impractical for the receiver to modify the response curve at every bit interval by physically tuning the receptor dissociation constant (RTAR) or expressing new receptors (REAR) due to the expected difference in timescales of data transmission and these biological receptor regulation mechanisms. Therefore, for tuning the response curve, we assume that receiver does not utilize its memory (i.e., $M=0$), and instead uses $\hat{\Omega}_0$ as the ISI estimate (which is independent of the symbol index $i$), such that the estimated shift is $\hat{\Delta} c_\mathrm{L} = \hat{\Omega}_0$. Then, using \eqref{eq:KDopt_shift}, we can write the optimal dissociation constant for the RTAR architecture as follows
\begin{align}
K_\mathrm{D,opt} = \sqrt{(c_{\mathrm{L}|0}^\ast + \hat{\Omega}_0)(c_{\mathrm{L}|1}^\ast + \hat{\Omega}_0)}.
\label{eq:KDopt_isi}
\end{align}
Similarly, for the REAR architecture, $K_\mathrm{D,opt}^\mathrm{new}$ is calculated through numerical optimization \eqref{eq:KDoptimization} using the BEP expression derived in Appendix \ref{AppendixA}.

The performance evaluation of the adaptive receiver architectures under ISI is conducted in two parts. First, we evaluate the performance as a function of the signaling interval, $T_\mathrm{S}$, which directly affects the level of interference. In this analysis, the receivers possess a memory of $2$ bits. In the baseline setting, $T_\mathrm{S}$ is assumed to be considerably large, thereby rendering ISI negligible for the receiver design. However, as $T_\mathrm{S}$ decreases, ISI becomes non-negligible and degrades detection performance. As shown in Fig. \ref{fig:BEP_isi_Ts}, the adaptive receiver architectures mitigate the impact of ISI on detection performance by nearly an order of magnitude when $T_\mathrm{S} \approx 4 \times t_\mathrm{Peak}$, despite continuing to exhibit a performance degradation trend as $T_\mathrm{S}$ decreases. 

Second, the receiver memory can significantly improve detection performance, as shown in Fig. \ref{fig:BEP_isi_memory}. This holds true even though the recently decoded bits stored in memory are not considered when modifying the response curve. The improvement stems from the optimized decision threshold, achieved through a more precise estimation of ISI by incorporating recently transmitted bits—which exert a higher impact—into the estimation scheme \cite{kuscu2018maximum}. 

\subsection{Case III: Stochastic Background Interference}
\label{sec:interference}

In the biological application environments of MC systems, various interference sources may exits, releasing the same type of ligands into the fluidic channel. This can lead to unintentional accumulation of ligand concentration at the receiver location, resulting in interference during detection. Such interference could originate from either natural biological sources generating the same ligands \cite{kuscu2022detection} or additional MC transmitters releasing identical molecules into the same channel (i.e., multi-user interference) \cite{dinc2017theoretical}. 

Adaptation against background concentration of ligands is a ubiquitous process in bacteria, allowing them to cope with varying environmental conditions and maintain chemotactic sensitivity. Bacteria enhance sensitivity by tuning receptor binding and activation rates through methylation and demethylation processes mediated by the enzymes CheR and CheB \cite{yi2000robust}. This adjustment enables them to adapt their response range (i.e., dynamic range) according to the ambient ligand concentration. In this way, bacteria can avoid receptor saturation or desensitization, and improve their signal detection and discrimination capabilities \cite{mello2007effects}.

In this analysis, we consider an environment with stochastic interference, where the interferer concentration at the receiver at each sampling time follows a log-normal distribution. We neglect ISI, assuming that $T_\mathrm{S}$ is sufficiently large. Additionally, we assume that the receiver is initially deployed with receptors featuring optimal $K_\mathrm{D}$ values for a baseline setting without interference. Consequently, when interference is introduced, the optimality of the receiver is compromised. This situation corresponds to the case where the received concentration signals are shifted with respect to the response curve, investigated in Section \ref{sec:shift}. However, in this practical scenario, the magnitude of the shift becomes a random variable rather than a constant value.

Considering stochastic interference, the received signal can be expressed as
\begin{align}
c_{\mathrm{L}|s} = c_{\mathrm{L}|s}^\ast + c_\mathrm{int},
\label{eq:received_signal_interference}
\end{align}
where $c_\mathrm{int}$ denotes the stochastic background interference, which follows a log-Normal distribution, i.e.,  $c_\mathrm{int} \sim \mathrm{Lognormal}\left(\mu_\mathrm{c_\mathrm{int}}, \sigma^2_\mathrm{c_\mathrm{int}}\right)$. As $c_\mathrm{int}$ is a random variable, the receiver cannot determine its exact value at the sampling time a priori, and thus, cannot utilize this information for tuning the response curve. To better understand the impact of this uncertainty, we investigate two scenarios with varying levels of the receiver's knowledge about the interference statistics. 
\subsubsection{Knowledge of the First Moment} In this scenario, the receiver has only the knowledge of the first moment, i.e., the mean, of the interferer concentration distribution, and considers the interference concentration as a deterministic variable. Therefore, the receiver's estimate of the received ligand concentration is given by 
\begin{align}
\hat{c}_{L|s} = c_{L|s}^\ast + \mu_\mathrm{c_\mathrm{int}}.
\label{eq:received_signal_interference_est}
\end{align}
The estimated concentration shift is then given by $\hat{\Delta} c_\mathrm{L} = \hat{\Omega}_0$. Utilizing \eqref{eq:KDopt_shift}, the optimal receptor dissociation constant for the RTAR architecture is determined as
\begin{align}
K_\mathrm{D,opt} = \sqrt{(c_{\mathrm{L}|0}^\ast + \mu_\mathrm{c_\mathrm{int}})(c_{\mathrm{L}|1}^\ast + \mu_\mathrm{c_\mathrm{int}}}).
\label{eq:KDopt_isi}
\end{align}
Likewise, for the REAR architecture, $K_\mathrm{D,opt}^\mathrm{new}$ can be obtained via numerical optimization \eqref{eq:KDoptimization} using the BEP expression \eqref{eq:bepbep}. 

\begin{figure}[!t]
	\centering
	\includegraphics[width=0.9\columnwidth]{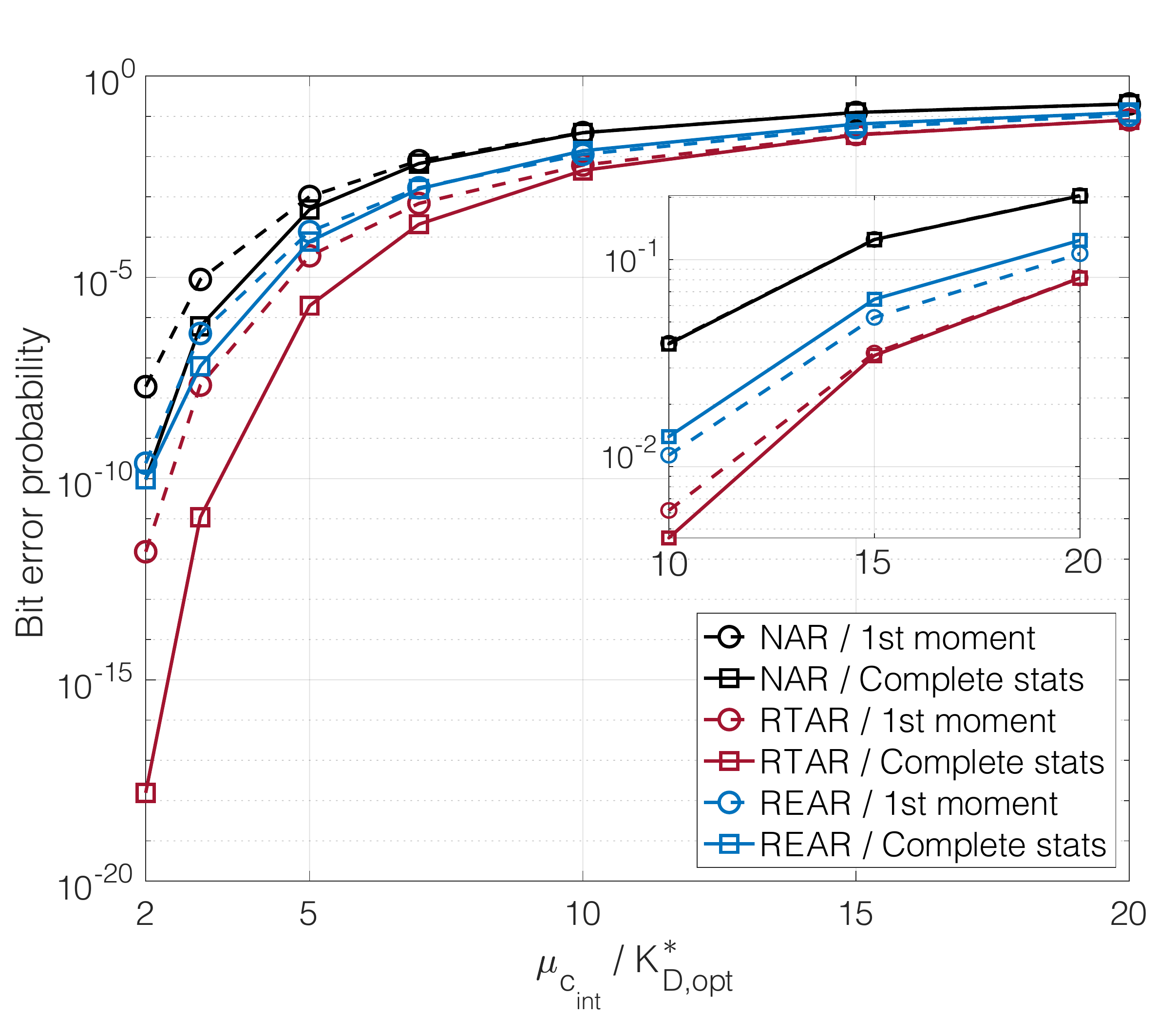}
	\caption{\emph{Adaptivity for stochastic interference:} BEP as a function of mean interference concentration for non-adaptive and adaptive receiver architectures with varying knowledge of interference statistics. }
    \label{fig:BEP_stoc_intf}
\end{figure}

\subsubsection{Complete Statistical Knowledge}
In this case, a complete statistical description of the interference is available to the receiver. By utilizing the probability distribution of the interferer concentration at the receiver, the optimal dissociation constant for the RTAR architecture can be determined numerically by solving the following optimization problem: 
\begin{align}
K_\mathrm{D,opt} = \argmin_{K_\mathrm{D}} \mathrm{BEP},
\label{eq:KDoptimizationSingleReceptor}
\end{align}
where BEP is given in \eqref{eq:bepbep}. To obtain BEP for this scenario, however, the mean and variance of the number of bound receptors, $N_\mathrm{B}$, corresponding to bit-$0$ and bit-$1$ transmissions must be calculated using the law of total expectation and the law of total variance, respectively, as explained in Appendix \ref{AppendixA}. Similarly, the optimal dissociation constant of the newly expressed receptors for the REAR architecture is obtained numerically by solving the optimization problem given in 
\eqref{eq:KDoptimization}. 

The results of the error performance analysis with varying mean interference concentration $\mu_\mathrm{c_{int}}$ is provided in Fig. \ref{fig:BEP_stoc_intf} for different receiver architectures. In this analysis, we assume that the standard deviation of the interference concentration takes the form of $\sigma_\mathrm{c_{int}} = 0.1 \times \mu_\mathrm{c_{int}} $. The results in terms of BEP reveal the same trend as in previous scenarios, with RTAR outperforming REAR, while both demonstrate superior performance compared to the non-adaptive architecture, NAR. It is evident that having complete knowledge of the interference statistics significantly aids in optimizing the receiver response curve. Note that the performance improvement in NAR, owing to complete statistical knowledge, stems exclusively from the more accurate optimization of the decision threshold. For the adaptive architectures, we observe that even knowledge of only the first moment can lead substantial improvements in detection performance, spanning several orders of magnitude. This improvement is more pronounced when the interference concentration has low to moderate expected values. 

\section{Conclusion}
\label{sec:conclusion}

In this study, we presented bio-inspired adaptive MC receiver architectures that can maintain optimal error performance in time-varying MC channels by modulating the response curve of their cell-surface sensory systems, consisting of ligand receptors. The performance of adaptive receivers was evaluated across a range of practical MC scenarios, encompassing three sources of time-variance in received signals, i.e., enzymatic degradation of information-carrying ligands, ISI, and stochastic background interference. Our analysis reveals a significant improvement in error performance with adaptivity in all investigated cases. We believe that this theoretical investigation is both timely and relevant, as it well aligns with recent advancements in synthetic biology, such as engineering and \emph{de novo} design of synthetic receptors with tunable binding characteristics, and synthetic chemical reaction networks capable of performing regulatory functions. Future research directions include developing time-resolved models to better understand the limitations and opportunities of adaptive MC receiver architectures, integrating dynamic tunability of ligand-receptor interactions into MC simulation frameworks, and developing higher-level adaptive network architectures for IoBNT, leveraging the tunability and adaptivity of biosynthetic MC devices. 

\appendices
\section{Derivation of BEP}
\label{AppendixA}

Utilizing the Gaussian approximation of the binomial distribution, we can express the conditional probability of observing $n_\mathrm{B}$ bound receptors at a specific sampling time, given that the transmitted bit is $s$, as follows:
\begin{align}
\p(n_\mathrm{B}|s) = \mathcal{N}\left(\E[n_\mathrm{B}|s],\Var[n_\mathrm{B}|s]\right),
\label{eq:nBmeanvariance}
\end{align}
where the mean and variance are obtained as
\begin{align}
\E[n_\mathrm{B}|s] &= \p_\mathrm{B}(c_\mathrm{L|s}) N_\mathrm{R} \\ \nonumber
\Var[n_\mathrm{B}|s] &= \p_\mathrm{B}(c_\mathrm{L|s}) \left(1-\p_\mathrm{B}(c_\mathrm{L|s})\right) N_\mathrm{R}.
\label{eq:nBmeanvariance}
\end{align}

In case of interference, the mean of $n_\mathrm{B}$ given the transmitted symbol can be obtained via the law of total expectation, i.e., 
\begin{align}
\E[n_\mathrm{B}|s] &= \E\Bigl[ \E[n_\mathrm{B} |s, c_\mathrm{int}] \Big| s\Bigr] \\ \nonumber
&= \int_0^\infty N_R \frac{c_{\mathrm{L}|s}^\ast + c_\mathrm{int}} {c_{\mathrm{L}|s}^\ast + c_\mathrm{int} + K_\mathrm{D}} \p(c_\mathrm{int}) dc_\mathrm{int},
\label{eq:nBmeanTotal}
\end{align}
where $\p(c_\mathrm{int}) = \mathrm{Lognormal}\left(\mu_\mathrm{c_\mathrm{int}}, \sigma^2_\mathrm{c_\mathrm{int}}\right)$. Likewise, the variance of $n_\mathrm{B}$ can be obtained using law of total variance as follows:
\begin{equation}
\Var[n_\mathrm{B}|s] = \E\Bigl[ \Var[n_\mathrm{B} |s, c_\mathrm{int}] \Big| s\Bigr] + \Var\Bigl[ \E[n_\mathrm{B} |s, c_\mathrm{int}] \Big| s\Bigr].
\label{eq:nBvarianceTotal}
\end{equation}
The explicit form of \eqref{eq:nBvarianceTotal} is not written for brevity; however, note that the involved integrals are calculated numerically. 

Considering that the system utilizes binary CSK, the decision rule can be expressed as
\begin{equation}
\hat{s} = \argmax_{s \in \{0,1\}} \p(n_\mathrm{B}|s),
\label{gaussian_DRBT2}
\end{equation}
where $\hat{s}$ is the decoded bit. Further simplification of the decision rule is achieved by defining a decision threshold $\lambda_{n_\mathrm{B}}$: 
\begin{equation} 
n_\mathrm{B} \underset{\hat{s} = 0}{\overset{\hat{s} = 1}{\gtrless}} \lambda_{n_\mathrm{B}}.
\label{threshold}
\end{equation}
For normally distributed statistics, the optimal decision threshold that minimizes the error probability is computed as
\begin{align} \label{threshold3}
 \lambda_{n_\mathrm{B}} &= \Gamma^{-1}  \Biggl( \Var[n_\mathrm{B}|1] \E[n_\mathrm{B}|0] \\ \nonumber
 &~~~ - \Var[n_\mathrm{B}|0] \E[n_\mathrm{B}|1]  + \Std[n_\mathrm{B}|1]\Std[n_\mathrm{B}|0] \\ \nonumber
&~~~ \times \sqrt{\bigl(\E[n_\mathrm{B}|1] - \E[n_\mathrm{B}|0] \bigr)^2+2 \Gamma  \ln{\frac{\Std[n_\mathrm{B}|1]}{\Std[n_\mathrm{B}|0]}}} ~\Biggr), 
\end{align}
where $\Gamma = \Var[n_\mathrm{B}|1] - \Var[n_\mathrm{B}|0]$, and $\Std[.] = \sqrt{\Var[.]}$ denotes the standard deviation. Given the decision thresholds and equal transmission probabilities, i.e., $\p_1 = \p_0 = 1/2$, BEP can be obtained as follows:
\begin{align} \label{eq:bepbep}
&\mathrm{BEP} =\frac{1}{2} \big[ \p(\hat{s} = 1|s=0) + \p(\hat{s} = 0|s=1) \big] \\ \nonumber
&=\frac{1}{4}\Biggl[\erfc \Biggl(\frac{\lambda_{n_\mathrm{B}} - \E[n_\mathrm{B}|0] }{\sqrt{2 \Var[n_\mathrm{B}|0] }}\Biggr) +  \erfc \Biggl(\frac {\E[n_\mathrm{B}|1]  - \lambda_{n_\mathrm{B}}}{\sqrt{2 \Var[n_\mathrm{B}|1]} }\Biggr)\Biggr],
\end{align}
where $\mathrm{erfc}(z)= \frac{2}{\sqrt{\pi}} \int_{z}^{\infty} e^{-y^2}dy$ is the complementary error function.

\bibliographystyle{IEEEtran}
\bibliography{references}

\end{document}